\documentclass[12pt,a4paper]{article}

\usepackage{amsmath}
\usepackage{amsfonts}
\usepackage{natbib}
\usepackage[singlespacing]{setspace}
\usepackage{authblk}
\usepackage{graphicx}
\usepackage[small]{titlesec}
\usepackage{geometry}
\usepackage{xcolor}
\usepackage{booktabs}
\usepackage{threeparttable}
\usepackage{mathtools}
\usepackage{hyperref}
\usepackage{lmodern}
\usepackage{endnotes}
\usepackage{appendix}

\let\footnote=\endnote

\providecommand{\keywords}[1]{\vspace{0.5em}\noindent\textbf{Keywords:} #1}

\title{\normalsize Mutual Information as a Tool for Optimal Classification: Application to Identifying Rapid-Responding Behaviour}
\author[1]{Santeri Holopainen}
\author[1]{Jari Metsämuuronen}
\author[1]{Mikko-Jussi Laakso}
\author[2]{Janne Kujala}
\affil[1]{Turku Research Institute for Learning Analytics, University of Turku}
\affil[2]{Department of Mathematics and Statistics, University of Turku}
\date{\small\today}

\begin{document}

\maketitle

\begin{abstract}
    \footnotesize Existing methods for identifying rapid-responding behaviour in large-scale assessments require parametric assumptions about the population. In this study, we propose a novel, non-parametric, mutual information-based framework of methods as an alternative. The methods within this framework compute the mutual information of the observed responses and discretised response times and maximise the information gain to determine a threshold that differentiates rapid responses from engaged responses. The only difference between the methods is the number of categories in the relevant variables. We present three methods explicitly. The first method uses response correctness and binarised response times. The second method uses correctness and categorises time into three groups. The third method uses raw responses and binarised times. We applied these methods to mathematics achievement data collected through the Programme for International Student Assessment in 2022. Furthermore, we examined the behaviour and usability of the first method in certain realistic conditions at the population and realised levels. The results indicated that the proposed framework is a viable alternative for identifying rapid responses. The framework's novelty lies in its non-parametric nature and its ability to utilise raw responses instead of correctness. Finally, we discuss some possible future research directions on this topic.
\end{abstract}


\newpage

\section{Introduction}

In large-scale assessment programmes designed to assess individuals' knowledge, skills, and abilities (KSAs), the phenomenon of \textit{test-taker disengagement} has attracted considerable interest~\citep[e.g.,][]{guo2024ai,rios2021meta,skalka2024,ulitzsch2023probabilistic}. This causes a response process in which a test taker does not show their true ability and hinders the validity of the observed test score~\citep[e.g.,][]{rios2022meta,wise2017rgb} and imposes bias to individual ability and item parameter estimates~\citep[e.g.,][]{guo2016cump,rios_2022a,rios_2022b,rios2024comparison,rios_soland_2021,wise_demars_2006}. Because of computer-based tests (CBTs), which allow the test administrator to measure the test takers’ \textit{response times} (RTs), disengagement research has focused on identifying \textit{rapid-responding behaviour}~\citep[RRB; or more traditionally, rapid-guessing behaviour; see][]{wise2017rgb}, a response process in which a test taker answers an item too quickly to have fully considered it. In applied research and operational settings, the most common approach to identifying RRB has been to use \textit{response time threshold methods}~\citep[RTTMs; e.g.,][]{guo2020differential,guo2016cump,holopainen_gauging_2025,holopainen2026,lee2014vitp,rios2020mln,schnipke1995speededness,wise2019information,wise2005rte,wise2012normative,wise2004investigation}. These methods use the observed responses and RTs in the data to estimate appropriate time thresholds that differentiate between rapid and engaged responses. 

To date, the most promising RTTMs are the \textit{Cumulative Proportion}~\citep[CUMP;][]{guo2016cump} and the \textit{Mixture Log-Normal}~\citep[MLN;][]{rios2020mln} methods, as they have the strongest theoretical justification and can be automatised~\citep[see, e.g.,][]{holopainen_gauging_2025,holopainen2026}, though empirical comparisons have provided mixed validation results~\citep[e.g.,][]{holopainen_gauging_2025,holopainen2026,krohne_2020,rios2021meta,rios2020mln,skalka2024,soland_2021}. However, these methods have some major limitations. First, they are limited by parametric assumptions about the underlying distributions. The CUMP method assumes that rapid responses are random, meaning the accuracy level of rapid responses equals the chance level, while the MLN method assumes that the RT distribution is a mixture of two log-normal distributions. Second, the CUMP method cannot be used when the responses cannot be evaluated in terms of their correctness (e.g., responses to questionnaires). Third, the MLN method cannot be used when the observed RT distribution is not bimodal. Finally, one of the major challenges in identifying RRB is the binary conception of test-taker disengagement as a psychological phenomenon when using RTTMs to differentiate between RRB and SB~\citep[see][]{holopainen_gauging_2025} and the question arises whether this limitation could be addressed somehow.

Based on the discussion above, we claim that disengagement researchers and practitioners would benefit from a methodology that could overcome the challenges in identifying RRB. For this purpose, we introduce a novel, non-parametric, mutual information-based framework of methods as an alternative tool for identifying RRB. Therefore, the remaining part of the paper proceeds as follows: First, we introduce mutual information generally, and then we bring it into the context of educational and psychological measurement as a measure of non-linear dependency between responses and discretised RTs. Second, we define the empirical counterpart of this measure, introduce the proposed MI-based framework for identifying RRB, and discuss its advantages compared to existing methods. Third, we illustrate the application of three MI-based methods to a subset of the PISA 2022 data. Fourth, we present the results of investigating one of the methods within an existing frame of assumptions when the population contains disengaged test takers at both the population and realised levels. Finally, we discuss possible future directions for this line of research.

\section{Mutual Information of Two Discrete Random Variables}

\textit{Mutual information} (MI) of two random variables is a measure of the amount of information obtained about one of the variables by observing the other variable \citep{cover1991elements}\footnote{The mathematical basis was established by~\citet{shannon_1948}, and the term was first coined by Robert Fano~\citep[see][]{kreer_1957}.}. Let $X$ with support $\mathcal{X}$ and $Y$ with support $\mathcal{Y}$ denote discrete random variables. Furthermore, let $\mathrm{P}_{X,Y}(x,y)$, $\mathrm{P}_{X}(x)$, and $\mathrm{P}_{Y}(y)$ denote $X$'s and $Y$'s joint probability mass distribution and marginal probability mass distributions, respectively. The MI of $X$ and $Y$ is defined as
\begin{equation} \label{eq_mi_def}
    \mathrm{I}(X,Y) = \sum_{x\in\mathcal{X}}\sum_{y\in\mathcal{Y}}\mathrm{P}_{X,Y}(x,y)\log_2\left(\frac{\mathrm{P}_{X,Y}(x,y)}{\mathrm{P}_{X}(x)\mathrm{P}_{Y}(y)}\right), 
\end{equation}
where the convention $0\log_2\frac{0}{z}=0$ is used for any non-zero real number $z$. That is, MI of $X$ and $Y$ evaluates the difference between their joint distribution and the product of their marginal distributions in some units of information, such as bits, as here, where a base-2 logarithm is used. The MI of $X$ and $Y$ can also be expressed in terms of the entropy of $Y$, $\mathrm{H}(Y)$, and the conditional entropy of $Y$ on $X$, $\mathrm{H}(Y \mid X)$, as follows:
\begin{equation}
    \mathrm{I}(X,Y) = \mathrm{H}(Y) - \mathrm{H}(Y \mid X),
    \label{eq_mi_def_v2}
\end{equation}
where 
\begin{equation}
    \mathrm{H}(Y) = -\sum_{y\in\mathcal{Y}}\mathrm{P}_{Y}(y)\log_2(\mathrm{P}_{Y}(y))
    \label{eq_entropy_of_Y}
\end{equation}
and
\begin{align}
    \mathrm{H}(Y \mid X) 
    &= \sum_{x\in\mathcal{X}}\mathrm{P}_{X}(x)\mathrm{H}(Y \mid X=x) \nonumber \\
    &= -\sum_{x\in\mathcal{X}}\mathrm{P}_{X}(x)\sum_{y\in\mathcal{Y}}\mathrm{P}_{Y \mid X}(y \mid x)\log_2(\mathrm{P}_{Y \mid X}(y \mid x)),
    \label{eq_conditional_entropy_of_Y_on_X}
\end{align}
where $\mathrm{P}_{Y \mid X}(y \mid x) = \mathrm{P}_{X,Y}(x,y) / \mathrm{P}_{X}(x)$ \citep{cover1991elements}.

In general, mutual information has the following properties: It is symmetric, that is, $\mathrm{I}(X,Y)=\mathrm{I}(Y,X)$, and it is always non-negative, that is, $\mathrm{I}(X,Y)\geq0$ for every $X$ and $Y$, and the equality holds if and only if $X$ and $Y$ are independent \citep{cover1991elements}. In the case of binary variables, the maximum MI is 1 and is reached when $\mathrm{P}_{X}(x)=\mathrm{P}_{Y}(y)=0.5$ and $X = Y$.

\section{Mutual Information as a Tool to Identify RRB}

MI has been widely used in a variety of applied research, such as signal processing~\citep[e.g.,][]{shannon_1948}, approximation of discrete probability distributions~\citep[e.g.,][]{chow_approximating_1968}, image registration~\citep[e.g.,][]{viola_alignment_1997}, and feature selection~\citep[e.g.,][]{peng_feature_2005}. MI has also been used in psychometric applications, such as item selection in computerised adaptive tests~\citep[e.g.,][]{wang_mutual_2013,weissman_mutual_2007} or Bayesian adaptive estimation of psychometric models~\citep[e.g.,][]{kujala_2006}. In the following, we bring MI into the context of educational and psychological measurement as a measure of non-linear dependency between responses and discretised RTs.

\subsection{MI of Responses and Discretised RTs}

Let us assume that we have a population of test takers answering the $j$th item of an educational or psychological test assessing some latent construct. Let $T_j$ with support $\mathcal{T} = \mathbb{R}^+$ and $Y_j$ with support $\mathcal{Y} = \{0, 1, ..., q-1\}$ denote the random variables for the RTs and responses on item $j$, respectively. More specifically, $Y_j$ is a general notation for responses and is either the random variable for the actual raw response to a multiple-choice question or the response's correctness and may take on $|\mathcal{Y}| = q > 1$ values. To avoid confusion, we shall denote responses generally with $Y_j$ and switch to $U_j$ when explicitly dealing with response correctness, for which $U_j = 1$ indicates a correct response and $U_j = 0$ an incorrect response. The marginal probability mass distributions of $Y_j$ and $U_j$ are denoted with $\mathrm{P}(Y_j = y)$, $y \in \mathcal{Y}$, and $\mathrm{P}(U_j = u)$, $u \in \{0,1\}$, respectively.

We define a discretised RT variable with support $\mathcal{X} = \{0, 1, ..., p-1\}$ that may take on $|\mathcal{X}| = p > 1$ values and is governed by $p - 1 > 0$ time thresholds as
\begin{equation}
    X_j = 
    \begin{cases}
        0,   \, \mathrm{if}\, T_j \leq t_0, \\
        1,   \, \mathrm{if}\, t_0 < T_j \leq t_1, \\
        \vdots \\
        p-2,   \, \mathrm{if}\, t_{p-3} < T_j \leq t_{p-2}, \\
        p-1,   \, \mathrm{if}\, T_j > t_{p-2},
    \end{cases}
\end{equation}
where $0 < t_0 < t_1 < \cdots < t_{p-3} < t_{p-2} < \infty$ are fixed time thresholds. Therefore, the marginal probability mass distribution of $X_j$ is defined as
\begin{equation}
    \mathrm{P}(X_j = x) =
    \begin{cases}
        \mathrm{P}(T_j \leq t_x), \, \mathrm{if}\, x = 0, \\
        \mathrm{P}(t_{x-1} < T_j \leq t_x), \, \mathrm{if}\, 0 < x < p-1, \\
        \mathrm{P}(T_j > t_{x-1}), \, \mathrm{if}\, x = p-1.
    \end{cases}
\end{equation}
In addition, the joint probability mass distribution of $X_j$ and $Y_j$ is
\begin{equation}
    \mathrm{P}(X_j = x, Y_j = y) = 
    \begin{cases}
        \mathrm{P}(T_j \leq t_x, Y_j = y), \, \mathrm{if}\, x = 0, \\
        \mathrm{P}(t_{x-1} < T_j \leq t_x, Y_j = y), \, \mathrm{if}\, 0 < x < p-1, \\
        \mathrm{P}(T_j > t_{x-1}, Y_j = y), \, \mathrm{if}\, x = p-1.
    \end{cases}
\end{equation}

Finally, the MI of $X_j$ and $Y_j$, i.e., the \textit{MI of Response and (discretised) Time} (MIRT), can be computed with
\begin{align}
    \mathrm{I}(X_j,Y_j) 
    & = \mathrm{H}(Y_j) - \sum_{x\in\mathcal{X}}\mathrm{P}(X_j = x)\mathrm{H}(Y_j \mid X_j=x) \nonumber \\
    & = \mathrm{H}(Y_j) \nonumber \\
    & \quad - \mathrm{P}(T_j \leq t_0)\mathrm{H}(Y_j \mid T_j \leq t_0) - \mathrm{P}(T_j > t_{p-2})\mathrm{H}(Y_j \mid T_j > t_{p-2}) \nonumber \\
    & \quad - \sum_{x = 1}^{p-2}\mathrm{P}(t_{x-1} < T_j \leq t_{x})\mathrm{H}(Y_j \mid t_{x-1} < T_j \leq t_{x}).
    \label{eq_mirt_def}
\end{align}
MIRT is a function of the time thresholds, and maximising MIRT maximises the information gained about the responses when the time thresholds are observed. In other words, maximising MIRT yields time thresholds that optimally classify the population into $p$ subgroups from an information-gain perspective. While this measure could be used to detect many different groups of test takers, our main focus is on finding a suitable threshold for identifying RRB. If the population is a mixture of RRB and SB, maximising MIRT could identify a suitable threshold for differentiating between the two groups, provided that the joint distributions of the responses and RTs are sufficiently different between them so that some other characteristic of the population is not captured as a maximum of MIRT instead. For example, the population could contain a shift from fast, low-performing test takers to slow, high-performing test takers as a function of RT~\citep[e.g., speed-accuracy trade-off; see][]{heitz_2014}.

MIRT may contain more than one local maximum as a function of the time threshold, and the global maximum may not be suitable for identifying RRB specifically. To overcome this limitation, the search space for the global maximum can be restricted to a suitable region where the RRB threshold is expected to lie, based on the common assumption that rapid responses are faster than engaged responses on average. The search space can then be limited to the interval $(0, t_{\mathrm{upp}}]$, where $t_{\mathrm{upp}}$ is a suitable time point. Next, we define empirical MIRT for finite samples at the realised level, which will be used in the the proposed MI-based framework of methods for RRB identification.

\subsection{Empirical MIRT}

Let us assume that we get a sample of $N$ independent observations, $(t_{1j}, ..., t_{Nj})$ from $T_j$ and $(y_{1j}, ..., y_{Nj})$ from $Y_j$. The marginal probabilities $\mathrm{P}(Y_j = y)$, $y \in \mathcal{Y}$, are estimated with the empirical probabilities
\begin{equation}
    \mathrm{P}_N(Y_j=y)=\frac{1}{N}\sum_{i=1}^N[y_{ij} = y],
\end{equation}
where $[y_{ij} = y]$ is the Iverson bracket of the statement $y_{ij} = y$ and equals $1$ if $y_{ij} = y$ is true and $0$ otherwise. To facilitate the computation of the empirical MIRT, we divide the observed RTs into $1 < p < N$ categories\footnote{Note that the number of $p-1$-tuples and, consequently, the computational demand increase rapidly as $p$ approaches half of the sample size. For example, if our sample size is 1,000; using two thresholds ($p=3$) produces $\binom{1000}{2}/1000 = 499.5$ times more tuples than using only one threshold ($p=2$), while using three thresholds ($p=4$) produces 166,167 times more tuples.}. Let $k$ index the $\binom{N}{p-1}$ increasing $p-1$-tuples of time thresholds. The $k$th set is defined as a subset of $(t_{1j}, ..., t_{Nj})$ as follows: 
\begin{equation}
    A_k = (t_{k_1}, ..., t_{k_{p-1}})  \subset (t_{1j}, ..., t_{Nj}), \, t_{k_1} < ... < t_{k_{p-1}}, \, 1 < p < N.
\end{equation}
For each $k \in \left\{1, ..., \binom{N}{p-1}\right\}$, we define $X_{kj}$ as
\begin{equation}
    X_{kj} = 
    \begin{cases}
        0,   \, \mathrm{if}\, T_j \leq t_{k_1}, \\
        1,   \, \mathrm{if}\, t_{k_1} < T_j \leq t_{k_2}, \\
        \vdots \\
        p-2,   \, \mathrm{if}\, t_{k_{p-2}} < T_j \leq t_{k_{p-1}}, \\
        p-1,   \, \mathrm{if}\, T_j > t_{k_{p-1}}
    \end{cases}
\end{equation}
and estimate the marginal probabilities $\mathrm{P}(X_{kj} = x)$, $x \in \mathcal{X}$, with the empirical probabilities
\begin{equation}
    \mathrm{P}_N(X_{kj} = x) = 
    \begin{cases}
        \frac{1}{N}\sum_{i = 1}^N[t_{ij} \leq t_{k_{x+1}}], \, \mathrm{if} \, x = 0, \\
        \frac{1}{N}\sum_{i = 1}^N[t_{k_{x}} < t_{ij} \leq t_{k_{x+1}}], \, \mathrm{if} \, 0 < x < p-1, \\
        \frac{1}{N}\sum_{i = 1}^N[t_{ij} > t_{k_{x}}], \, \mathrm{if} \, x = p-1.
    \end{cases}
\end{equation}
Furthermore, we estimate the joint probabilities $\mathrm{P}(X_{kj} = x, Y_j = y)$, $x \in \mathcal{X}$, $y \in \mathcal{Y}$, with the empirical joint probabilities 
\begin{equation}
    \mathrm{P}_N(X_{kj} = x, Y_j = y) = 
    \begin{cases}
        \frac{1}{N}\sum_{i = 1}^N[t_{ij} \leq t_{k_{x+1}} \cap y_{ij} = y], \, \mathrm{if} \, x = 0, \\
        \frac{1}{N}\sum_{i = 1}^N[t_{k_{x}} < t_{ij} \leq t_{k_{x+1}} \cap y_{ij} = y], \, \mathrm{if} \, 0 < x < p-1, \\
        \frac{1}{N}\sum_{i = 1}^N[t_{ij} > t_{k_{x}} \cap y_{ij} = y], \, \mathrm{if} \, x = p-1.
    \end{cases}
\end{equation}
Finally, for each $k$, the empirical MI of $X_{kj}$ and $Y_j$, i.e., the empirical MIRT, is defined as
\begin{equation}
    \mathrm{I}_N(X_{kj},Y_j) = \mathrm{H}_N(Y_j) - \sum_{x\in\mathcal{X}}\mathrm{P}_N(X_{kj} = x)\mathrm{H}_N(Y_j \mid X_{kj}=x),
\end{equation}
where $\mathrm{H}_N(\cdot)$ is the empirical entropy and is computed using the corresponding empirical probability.

\subsection{Framework of MI-based Methods for RRB Identification}

Instead of proposing a single method for identifying RRB, we present a framework of MI-based methods, known as the \textit{Maximal Mutual Information} (MaxMI) framework. These methods are based on the change of the empirical MIRT as a function of time thresholds and locating the global maximum of the empirical MIRT within a predefined search space. In the following, we introduce three MaxMI methods that differ in the number of categories in $X_j$ and $Y_j$. The methods are abbreviated as "MaxMI-PQ", where "P" and "Q" are replaced with the corresponding number of categories in $X_j$ and $Y_j$, respectively.

\subsubsection{MaxMI when Both Variables are Binary (MaxMI-22)}

The first MaxMI method presented is perhaps the most intuitive. It assumes that both $X_j$ and $Y_j$ are binary (i.e., $p = q = 2$), with $X_j$ governed by a single time threshold and $Y_j = U_j$ representing the random variable for response correctness\footnote{Note that when $Y_j$ is binary, it can also represent a random variable for a raw response to a two-choice item. In practice, however, it does not matter whether raw responses or correctness indicator are used, though treating $Y_j$ as the random variable for response correctness ($U_j$) facilitates using the method for all types of items (e.g., multiple-choice or open-response) instead of just two-choice items.}. In this special case, each $p-1$-tuple of time thresholds contains a single threshold. The search space for the threshold can be limited by making use of the properties of MI in the case of binary variables: $\mathrm{I}(X_{j},U_j) = 0$ if $\mathrm{P}(U_j=1\mid X_{j}=1) = \mathrm{P}(U_j=1\mid X_{j}=0)$\footnote{Similarly, $\mathrm{I}(X_{j},U_j) = 0$ if $\mathrm{P}(X_{j}=1, U_j = 1) = \mathrm{P}(X_j = 1)\mathrm{P}(U_j = 1)$.}, meaning the first local maximum of MIRT is located between 0 and first the intersection of $\mathrm{P}(U_j=1\mid X_{j}=1)$ and $\mathrm{P}(U_j=1\mid X_{j}=0)$. Therefore, the upper limit of the search space can be set to the smallest time point where $\mathrm{P}_N(U_j=1\mid X_{ij}=1) - \mathrm{P}_N(U_j=1\mid X_{ij}=0)$ switches its sign\footnote{More precisely, the upper limit will be some time point between the last time point at which the sign is positive (negative) and the first time point at which the sign is negative (positive).}. The MaxMI-22 threshold for identifying RRB is defined as the time point where the empirical MIRT is maximised within the predefined search space.

\subsubsection{MaxMI when the Discretised RT is Ternary and the Response is Binary (MaxMI-32)}

In the second MaxMI method, $X_j$ is assumed to be governed by two time thresholds and $Y_j=U_j$ is assumed again to denote the random variable for response correctness (i.e., $p = 3$ and $q = 2$). This is a natural extension of the first method because the distribution may contain more than two sub-groups of test takers, rather than just those exhibiting RRB and SB. For example, RRB test takers could be categorised as either answering very quickly without considering the questions, or quickly glancing at the questions before deciding to guess~\citep[see, e.g.,][]{wise2017rgb}. SB test takers could be categorised as either fast high performers or slow low performers, or fast low performers or slow high performers. The latter grouping would align with the speed-accuracy trade-off. Using MaxMI-32, two thresholds are obtained by maximising the empirical MIRT. In this case, it is not necessarily true that $\mathrm{I}(X_{j},U_j) = 0$ if $\mathrm{P}(U_j=1\mid X_{j}=x_1) = \mathrm{P}(U_j=1\mid X_{j}=x_2)$ for some $x_1 \neq x_2, x_1,x_2\in \{0,1,2\}$, meaning the search space cannot be restricted as meaningfully as with the MaxMI-22 method. It is the task of the data analyst to determine which threshold is more suitable for identifying RRB. As a general rule, the smaller of the two time thresholds could be chosen as the RRB threshold.

\subsubsection{MaxMI when the Discretised RT is Binary and the Response is the Raw Response (MaxMI-2Q)}

The third method is based on $X_j$ being binary and using the raw response variable rather than the variable indicating correctness. This means that $Y_j$ could be a nominal variable with more than two categories ($q$). Indeed, $Y_j$ could represent the raw response in a survey assessing the participants' attitudes, beliefs or opinions, meaning we are not limited to identifying disengaged responses in achievement tests. This is a natural extension of the first method because, if little information about response correctness is gained from observing response time, MaxMI-22 might not work. This could occur, for example, when the response accuracies of RRB and SB test takers are similar, or when the response accuracies of RRB test takers and the fastest SB test takers are similar and the direction of the association between ability and speed is negative. This would mean that the response times of low-performing SB test takers are similar to those of RRB test takers. However, using the raw response rather than correctness of the response could be useful, particularly if rapid responders are prone to selecting one of the options due to edge aversion~\citep[see][]{attali2003,wise_kuhfeld_2020} or some other bias~\citep[e.g.,][]{holopainen_gauging_2025}.

Similar to the MaxMI-32 method, it may not hold that $\mathrm{I}(X_{j},Y_j) = 0$ if $\mathrm{P}(Y_j=y\mid X_{j}=1) = \mathrm{P}(Y_j=y\mid X_{j}=0)$ for some $y\in \{0,...,q-1\}$. Therefore, when using MaxMI-2Q, the RRB threshold is simply defined as the time point at which the empirical MIRT is maximised globally. It should be noted that this method could be further extended by allowing $X_j$ to be governed by more than one time threshold.

\subsubsection{Advantages of the MaxMI Framework}

Let us briefly discuss the advantages of the MaxMI framework compared to existing methods. First, the framework provides an even stronger theoretical justification than the CUMP and MLN methods, as the MaxMI methods are used to split the data optimally from an information-gain perspective. Second, in the introduction of this study, we mentioned that the CUMP and MLN methods make parametric assumptions about the population and cannot be used in certain situations. However, the MaxMI framework is non-parametric and can be used even if only the raw responses that may have a nominal or ordinal scale are available (e.g., MaxMI-2Q). It can also be used even if the observed RT distribution is not bimodal, provided that the population truly contains RRB and SB and the responses have enough information to exhibit the differences between the two groups. Third, we also noted that one of the major challenges in identifying RRB is the binary conception of disengagement when using time thresholds to differentiate between RRB and SB. While this challenge is also relevant to the MaxMI methods, the framework provides an alternative way to approach the RRB identification problem, as some of the MaxMI methods produce more than one time threshold (e.g., MaxMI-32).

In general, the MaxMI methods can be used to compute thresholds as long as the empirical MIRT can be calculated for each $p-1$-tuple. The only problem arises when either $X_j$ or $Y_j$ is deterministic, i.e., when $\mathrm{P}_N(X_{kj} = x) = 1$ for some $x$, or $\mathrm{P}_N(Y_j = y) = 1$ for some $y$. However, $\mathrm{P}_N(X_{kj} = x) \neq 1$ for all $x$ with the given assumptions. In addition, assuming stochasticity in the responses produced by a sufficiently large sample of participants, $\mathrm{P}_N(Y_j = y) \neq 1$ for all $y$. However, if the sample size is small, say it contains fewer than 100 participants, all participants may answer correctly or incorrectly to a test item (meaning that all participants selected the same response option), in which case all RTTMs that utilise the responses fail to estimate a threshold.

\section{Illustrating the MaxMI Framework in PISA 2022 Data} \label{seq:empirical_example_pisa2022}

\subsection{Participants and Data}

To demonstrate how the three MaxMI methods can be used in practice to identify RRB, we applied them to data obtained from 98 multiple-choice items in the computer-based mathematics section of the PISA 2022 main survey\footnote{The data is available at the OECD website: https://www.oecd.org/en/data/datasets/pisa-2022-database.html}. A total of 81 countries participated in PISA in 2022, but we focused on the Finnish sample of 10,256 students. Not all students complete the same items in PISA. In our case, the number of students varied between 856 and 1,584 across the 98 items. Because PISA assessments are low stakes from the students' perspective, we expected RRB to stem mainly from test takers who were unmotivated to answer the items carefully. Further details about the testing and sampling design of PISA 2022 can be found in the technical report~\citep{oecd2024technical}.

\subsection{Results}

We used the R 4.5.0 software~\citep{r_core_team} for analysing the data. In addition to the MaxMI methods, we applied the \textit{Cumulative Proportion}~\citep[CUMP;][]{guo2016cump} and the \textit{Mixture Log-Normal}~\citep[MLN;][]{rios2020mln} methods to the PISA 2022 data. In short, the CUMP method assumes that rapid responses are random and defines the threshold as the last time point at which the curve of the cumulative proportion of correct responses intersects the chance level. The MLN method assumes that the underlying RT distribution is a mixture of two log-normal distributions and fits the assumed model to the observed RT data. The MLN threshold is defined as the time point where the estimated RT density is minimised between the estimated RT modes. See Appendix~\ref{app_CUMP_and_MLN_in_PISA_and_simulation} for the detailed description of the operational rules used to apply these methods to the PISA data.

\begin{figure}[t]
\centering
\includegraphics[width=0.8\textwidth]{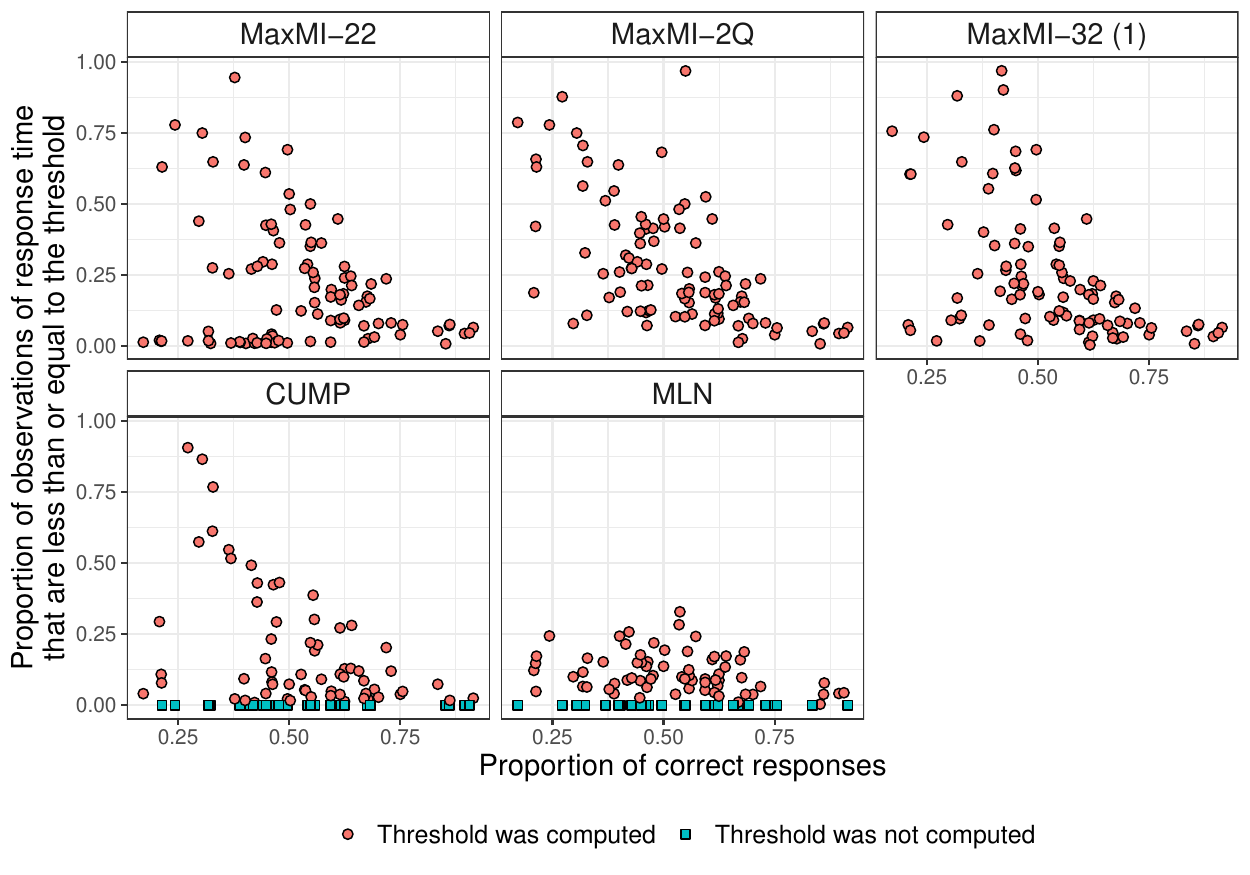}
\caption{Proportion of observations of response time that are less than or equal to the threshold as a function of the proportion of correct responses. Note: The proportion of observations was set to zero for those item-threshold pairs for which a threshold was not computed by CUMP or MLN. For the MaxMI-32 method, the plotted values are based on the first threshold}
\label{fig1}
\end{figure}

First, we analysed the thresholds of all 98 items. The CUMP and MLN methods were unable to compute thresholds for all of them, achieving 64 (65.3\%) and 68 (69.4\%) computed thresholds, respectively\footnote{Note that the result for the MLN method could vary if the threshold computation were repeated. This is because we let the starting values for the parameters to be randomly generated in the expectation-maximisation algorithm used in the 'mixtools' R package~\citep{benaglia2009} when fitting a two-component mixture model to the data.}. Figure~\ref{fig1} shows, for each method, the proportion of response time observations that are less than or equal to the threshold as a function of the proportion of correct responses. Excluding the values in the bottom-left corners of the plots, it appears that the proportion of observations for the MaxMI and CUMP methods decreases as the proportion of correct responses increases and diverges from the chance level ($=0.25$; most of the items were four-choice items). This does not apply to MLN, however, as it is unaffected by the response variable.

Second, we selected three four-choice items for further analysis to illustrate the behaviour of the MaxMI methods more clearly. These items were chosen for their ability to demonstrate the strengths and weaknesses of the methods. For each item, one MaxMI method successfully estimated a suitable threshold, while the others did not. Table~\ref{table1} presents the basic statistics and RRB thresholds for each of the three items. 

\begin{table}[t]
\begin{center}
\caption{Basic Statistics of the Finnish Sample and RRB Thresholds for Each of the Three PISA 2022 Mathematics Items\label{table1}}
\begin{tabular}{l c c c c c c c c c c}
\hline
     & \multicolumn{4}{c}{Sample statistics} & \multicolumn{6}{c}{RRB thresholds (sec.)} \\
Item & N & $P^{+}$ & RT Q1 & RT Md & A & B & C (1) & C (2) & D & E \\ 
\hline
MA131Q02 & 1,282 & 0.32 & 17 & 39 & \textbf{5} & 45 & 132 & 145 & - & \textbf{6} \\ 
MA144Q01 & 1,304 & 0.30 & 34 & 58 & 52 & \textbf{14} & 51 & 52 & 66 & \textbf{17} \\ 
MA147Q04 & 1,159 & 0.30 & 15 & 40 & 102 & 102 & \textbf{7} & 102 & 159 & - \\
\hline
\end{tabular}
\begin{tablenotes}
\item Note: N = sample size; $P^{+}$ = proportion of correct responses; RT Q1 = first RT quartile; RT Md = RT median; A = MaxMI-22; B = MaxMI-2Q; C = MaxMI-32, where (1) and (2) denote the first and second threshold, respectively; D = CUMP; E = MLN. Thresholds smaller than the first RT quartile are bolded.
\end{tablenotes}
\end{center}
\end{table}

\begin{figure}[!htbp]
\centering
\includegraphics[width=0.8\textwidth]{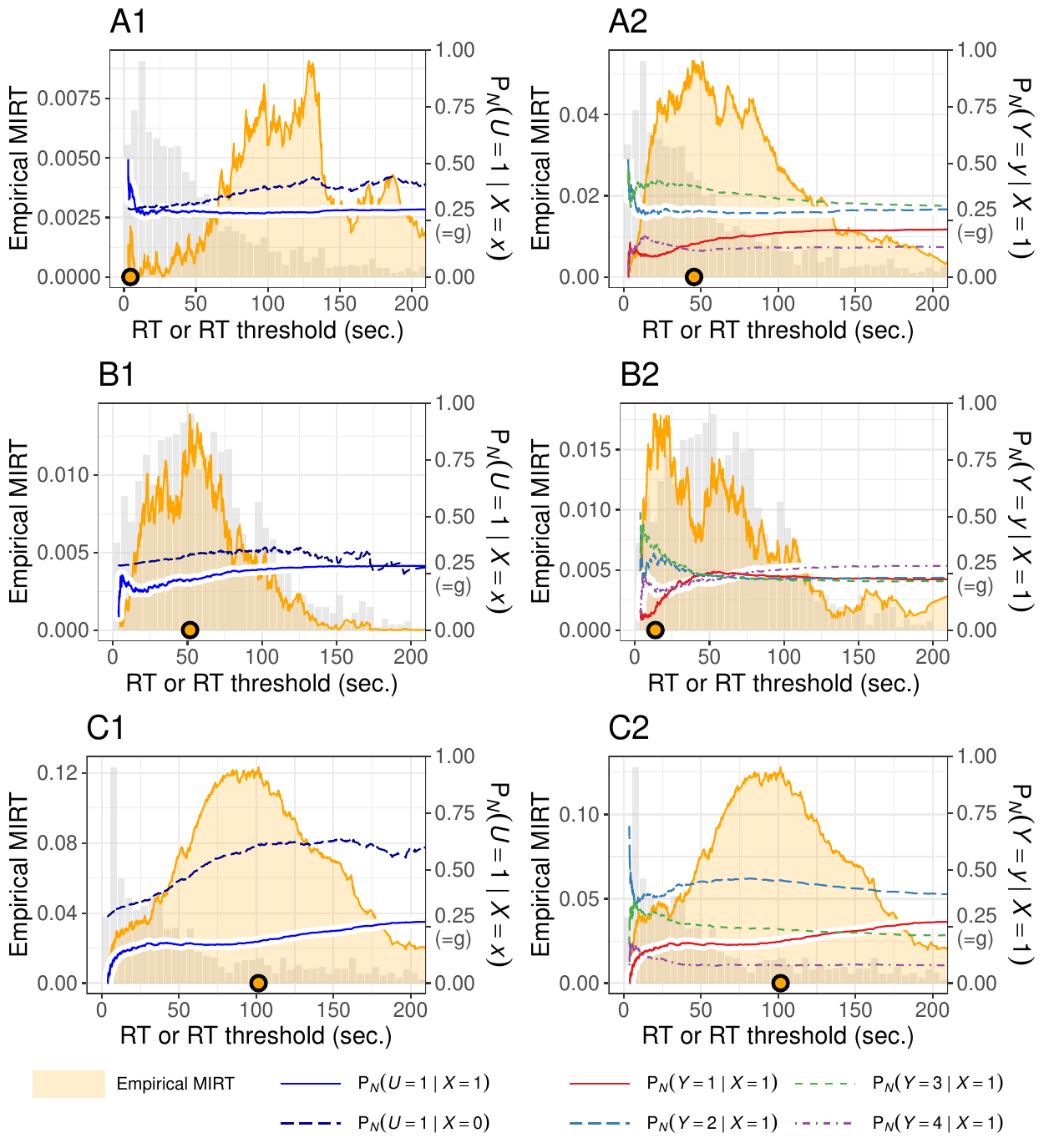}
\caption{Empirical MIRT and conditional response probabilities given the time threshold when the discretised RT variable is binary and the response variable denotes either correctness (plots A1, B1 and C1) or the raw response (plots A2, B2 and C2), for each of the selected items (MA131Q02 in plots A1 and A2; MA144Q01 in plots B1 and B2; and MA147Q04 in plots C1 and C2). Note: The threshold produced by the corresponding MaxMI is marked with a circle on the plots (plots A1, B1 and C1: MaxMI-22; plots A2, B2 and and C2: MaxMI-2Q). The observed RT distribution is illustrated as a histogram in the background. In plots A2, B2, and C2, the $\mathrm{P}_N(Y=y\mid X =1)$ that matches $\mathrm{P}_N(U=1\mid X = 1)$ in plots A1, B1, or C1, respectively, is highlighted with a white band. To reduce spurious noise in the curves, the ten observations with the smallest RTs were omitted}
\label{fig2}
\end{figure}

The only MaxMI method that produced a suitable threshold for the first item in Table~\ref{table1} (MA131Q02) was MaxMI-22, as the others were all greater than the median RT. Plots A1 and A2 in Figure~\ref{fig2} show the results for item MA131Q02. First, the observed RT distribution did not exhibit bimodality. Instead, most of the RTs accumulated relatively close to zero, forming one mode. This generally makes it difficult to find suitable thresholds for RRB identification because, in these cases, either the RTs of RRB test takers and most of the SB test takers are close to each other, or the RT distribution is not a mixture of RRB and SB. However, plot A1 shows that the cumulative proportion of correct responses \{$\mathrm{P}_N(U=1\mid X=1)$\} decreases sharply when moving from 0 to around 10 seconds, after which it remains stable. This type of behaviour indicates a mixture distribution. Interestingly, despite the unimodality of the observed RT distribution, MLN managed to compute a viable threshold. Second, the MaxMI-22 method benefited from limiting the search space for the global maximum of the empirical MIRT to be between 0 and the first intersection of $\mathrm{P}_N(U=1\mid X=1)$ and $\mathrm{P}_N(U=1\mid X=0)$. Without restricting the search space, the local maximum below the intersection would have remained unnoticed, as it was not the global maximum.

The only MaxMI method that computed a viable threshold for the second item in Table~\ref{table1} (MA144Q01) was MaxMI-2Q. Plots B1 and B2 in Figure~\ref{fig2} illustrate the results. First, it was debatable whether the observed RT distribution was bimodal. Clearly, one mode was located at around 50 seconds, but it could be argued that there was another mode at around 10, 60 and/or 100 seconds. Furthermore, in plot A1, the curve $\mathrm{P}_N(U=1\mid X=1)$ initially increased sharply. After this, it decreased for a few time segments before finally converging towards the overall proportion of correct responses. These findings suggest that the distribution may have contained two or more components. The behaviour of $\mathrm{P}_N(U=1\mid X=1)$ in particular suggests an RRB component accompanied by two SB subcomponents that behave differently, one of which contains fast test takers who answer correctly with lower accuracy than slower test takers (speed-accuracy trade-off). Similar to the first item, MLN produced a viable threshold despite the shape of the distribution. Second, limiting the search space of the global maximum of the empirical MIRT did not benefit MaxMI-22: the intersection of $\mathrm{P}_N(U=1\mid X=1)$ and $\mathrm{P}_N(U=1\mid X=0)$ was greater than the location of the global maximum of the empirical MIRT. In this case, the global maximum was at a time point that was not suitable for identifying RRB.

Finally, assuming that the response options 1--4 were presented in order, we can deduce that RRB test takers may have exhibited edge aversion, because the cumulative proportions of the second and third response categories (middle options) at small RTs (e.g., below 25 seconds) were much higher than those of the first or last (edge options). However, as RT increased, the proportions of the middle options decreased, while those of the edge options increased\footnote{The same phenomenon could be observed in the first item in plot A2, but it was less obvious because the correct option was one of the middle options, and the cumulative proportions of the middle options remained higher than those of the edge options.}. This phenomenon may have enabled MaxMI-2Q to estimate a suitable threshold, given that the cumulative response probabilities increased or decreased relatively quickly in relation to RT, which was captured by the empirical MIRT as a sharp increase at small RTs.

\begin{figure}[t]
\centering
\includegraphics[width=0.6\textwidth]{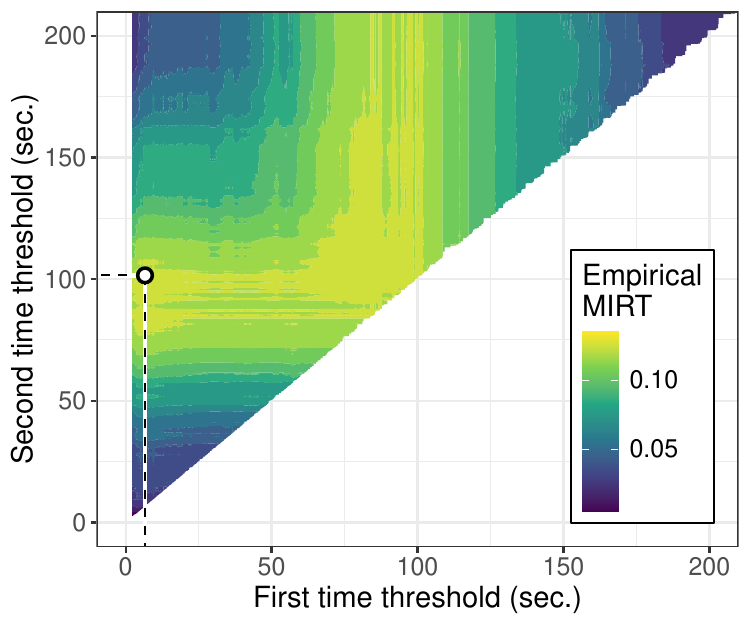}
\caption{Contour plot of the empirical MIRT when the discretised RT variable is governed by two time thresholds and the response variable denotes correctness for item MA147Q04. Note: The location of the global maximum of the empirical MIRT is marked on the plot with a circle}
\label{fig3}
\end{figure}

For the third and final item in Table~\ref{table1} (MA147Q04), only the MaxMI-32 method managed to compute an appropriate threshold. Plots C1 and C2 in Figure~\ref{fig2}, together with Figure~\ref{fig3}, illustrate the results. Figure~\ref{fig2} provides similar insights to those seen above: The RT distribution did not exhibit clear multimodality, but the $\mathrm{P}_N(U=1\mid X = 1)$ curve indicated two or more components. Limiting the search space of the global maximum did not benefit the MaxMI-22 method, and RRB test takers avoided edge options. However, Figure~\ref{fig3} adds to Figure~\ref{fig2} by showing that, when the empirical MIRT was governed by two time thresholds, it captured the sharp increase in $\mathrm{P}_N(U=1\mid X=1)$ when the first time threshold was kept fixed at approximately seven seconds.

In summary, these results suggest that the three MaxMI methods have different strengths and weaknesses, and can be applied to real-world data. The MaxMI-22 method is arguably be the most generalisable for identifying RRB, as it can be applied to both multiple-choice and open-response items, and it aligns with the existing RTTMs by producing only one threshold. Another benefit is that the search space can be narrowed down in a meaningful way. However, other methods within the framework may be advantageous in specific situations. For example, MaxMI-32 may be useful when RRB and/or SB test takers can be divided into smaller sub-groups, and MaxMI-2Q may be useful when the conditional marginal response probabilities of RRB and SB test takers are close to each other.

\section{Investigation of the Behaviour and Usability of MaxMI in the ILC-IRT Framework}

The \textit{Independent Latent Class Item Response Theory} (ILC-IRT) framework for response engagement incorporates RTs and treats them as indicators of response engagement. The framework implements a dichotomous latent class variable, $\Delta$, which varies within individuals and indicates whether a test taker is engaged with an item or not. This latent variable is assumed to influence both responses and RTs~\citep{nagy2022multilevel}. In other words, the framework assumes that the distributions of responses and RTs are mixture distributions, with one component for engaged test takers (SB; $\Delta = 1$) and one for disengaged test takers (RRB; $\Delta = 0$). For our purposes, ILC-IRT provides a means of examining the behaviour and usability of MaxMI when the population containing disengaged test takers is characterised by an existing, well-grounded set of assumptions.

In Appendix~\ref{app_ILC-IRT}, we provide a closer introduction to ILC-IRT. Here, we briefly describe the parameterisations used in this study, which follow the model in~\citet{ulitzsch2020hierarchical}. The person parameters---engagement tendency ($\phi$), ability ($\theta$), and speed ($\tau$)---were multivariate normally distributed. The probability of responding in an engaged state (SB component) was modelled using the one-parameter logistic (1PL) IRT model, where the person parameter was $\phi$ and the item parameter represented the engagement difficulty of the item ($\iota$), with larger values eliciting greater disengagement. In the SB component, the probability of answering correctly was also modelled using the 1PL IRT model, where the person parameter was $\theta$ and the item parameter represented the item's difficulty ($b$), whereas in the RRB component, the response probability equalled the chance level ($g$). Finally, in both components, the RTs were log-normally distributed, but with different parameterisations. In the SB component, the log-normal distribution contained person and item parameters, where the person parameter was $\tau$ and the item parameters specified the time intensity ($\beta$) and discrimination ($\alpha$) of the item \{$(\log(T) \mid \Delta = 1) \sim N(\beta-\tau,\alpha^2)$\}, whereas in the RRB component, we assumed a common mean and variance across items and test takers \{$(\log(T) \mid \Delta = 0) \sim N(\mu_c,\sigma_c^2)$\}.

The following investigation examines the behaviour of MaxMI at the population level and its usability for identifying RRB at the realised level under certain conditions according to the ILC-IRT framework. Of the MaxMI methods presented, the version with binary variables (MaxMI-22) was chosen for these investigations because the focus is on response correctness and the population is assumed to contain two explicit sub-groups: RRB and SB. We used the R 4.5.0 software~\citep{r_core_team} for numerical integration, as well as for generating and analysing the data. In multivariate numerical integration, the R package 'cubature'~\citep{cubature_package} was utilized.

\subsection{Population Level}

The examined conditions varied in terms of the direction of the association between ability and speed for SB test takers (negative or positive, i.e., $\rho_{\theta\tau} < 0$ or $\rho_{\theta\tau} > 0$)\footnote{Negative association is consistent with~\citet{ulitzsch2020hierarchical} and corresponds to the speed-accuracy trade-off. Positive association was added as an additional condition to reflect situations in large-scale assessments where the correlation between ability and speed is positive~\citep[e.g.,][]{holopainen2026}.}; the percentage of RRB in the population (5\% or 10\%)\footnote{The smaller RRB percentage represents items that are very easy to be engaged with, e.g., items at the beginning of a test or items that do not require much reading. The larger percentage represents the average RRB proportion of 10\% in low-stakes assessments found by~\citet{rios2022meta}.}; the RT mode for SB test takers (22 sec. or 60 sec.)\footnote{These correspond to the simulation study in~\citet{ulitzsch2020hierarchical}.}; and the marginal probability of a correct response for SB test takers (100 distinct values between 0.029 and 0.974). Therefore, the total number of different conditions was $2\times 2 \times 2 \times 100 = 800$. Table~\ref{table2} displays the varied population characteristics and the corresponding ILC-IRT parameters. All other parameters were fixed to certain values. For example, the association of engagement tendency with ability and speed were assumed to be positive and all of the examined items were four-choice, meaning that $g = 0.25$. The full parameter selection for these investigations, along with its correspondence to earlier research, is described in Appendix~\ref{app_parameter_selections}.

\begin{table}[b]
    \begin{center}
    \caption{Varied Population Characteristics and the Corresponding Parameters for Investigating the Behaviour and Usability of MaxMI-22 in the ILC-IRT Framework \label{table2}}
        \begin{tabular}{@{}lll@{}}
        \hline
        Population characteristic & Values & ILC-IRT parameter \\
        \hline
        Direction of association between & Negative; positive & $\rho_{\theta\tau} \in \{-0.4, 0.4\}$ \\
        \quad ability and speed for SB test-takers & & \\
        Percentage of RRB & 5\%, 10\% & $\iota \in \{-4.27, -3.29\}$ \\
        RT mode for SB test-takers & 22 sec., 60 sec. & $\beta \in \{3.25, 4.25\}$ \\
        Conditional marginal probability & 100 values between & $b = -4 + \frac{8i}{99}$, \\
        \quad of a correct response for SB test-takers & \quad 0.029 and 0.974 & \quad $i \in \{0,...,99\}$ \\
        \hline
        \end{tabular}
        \begin{tablenotes}
            \item Note: $\rho_{\theta\tau}$ = correlation between ability speed; $\iota$ = engagement difficulty; $\beta$ = time intensity; $b$ = item difficulty. The total number of conditions was $2\times 2 \times 2 \times 100 = 800$.
        \end{tablenotes}
    \end{center}
\end{table}

\begin{figure}[!t]
\centering
\includegraphics[width=0.8\textwidth]{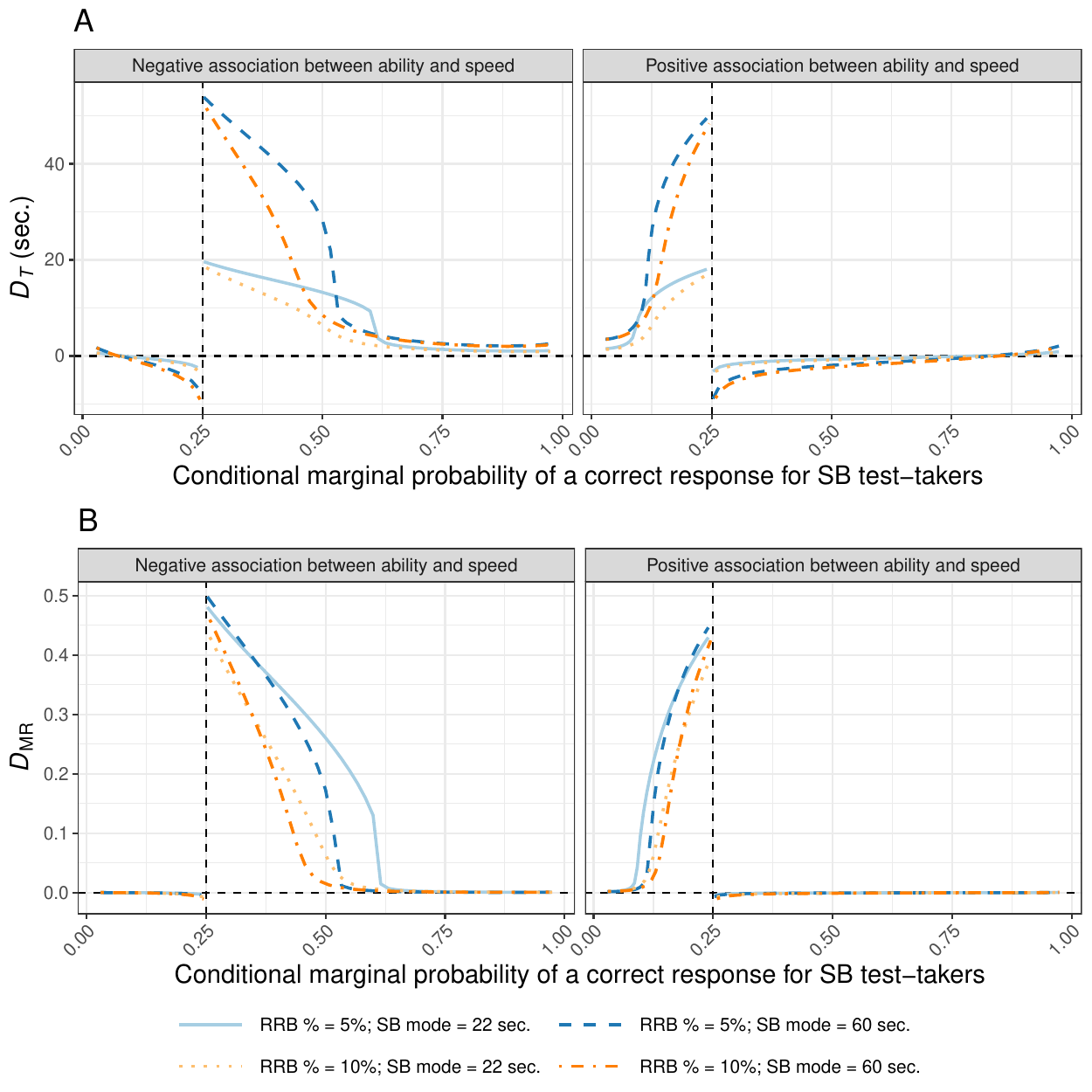}
\caption{Differences between (A) the thresholds ($D_T$; in seconds) and (B) the misclassification rates ($D_{\mathrm{MR}}$) produced by the MaxMI-22 method and by minimising the misclassification rate (MR) as functions of the marginal probability of a correct response for SB test takers. Note: RRB \% = RRB percentage; SB mode = RT mode for SB test takers. A negative difference between thresholds ($D_T < 0$) indicates that the MaxMI-22 threshold is smaller than the threshold that minimises the MR. The sign of $D_{\mathrm{MR}}$ was set to negative if the corresponding $D_T < 0$}
\label{fig4}
\end{figure}

We compared the time thresholds and misclassification rates (MRs) produced by the MaxMI-22 method with those obtained by minimising the MR. See the Online Resource of this article for the formulation of the MR in the ILC-IRT framework. Specifically, we examined the differences in thresholds ($D_{T}$) and misclassification rates ($D_{\mathrm{MR}}$) as functions of the conditional marginal probability of a correct response for SB test takers in groups categorised by the direction of the association between ability and speed, RRB percentage, and RT mode for SB test takers. Figure~\ref{fig4} plots these differences and summarises the results in two main findings. First, plot A shows an interaction effect between the direction of the association and the response probability on $D_T$. For cases with $\rho_{\theta\tau} < 0$, $D_T$ increased as the response probability approached the chance level $g=0.25$ from the right, and $D_T$ decreased as the response probability approached $g$ from the left, regardless of the RRB percentage or the RT mode. For cases with $\rho_{\theta\tau} > 0$, the same applied, but in reverse: $D_T$ decreased as the response probability approached $g$ from the right, and $D_T$ increased as the response probability approached $g$ from the left, regardless of the other characteristics. In terms of $D_{\mathrm{MR}}$, however, plot B shows that MaxMI-22 produced a substantial MR only when the MaxMI-22 threshold was much larger than the threshold that minimises MR. In other words, when the MaxMI-22 thresholds were smaller than the thresholds that minimised MR, the differences in MR were negligible.

Second, the differences in $D_T$ between groups categorised by RRB percentage and SB mode were evident only when the response probability was close to $g$. Interestingly, for cases with $\rho_{\theta\tau} < 0$, the $D_T$ value for the group with an RRB percentage of 5\% and an SB mode of 22 seconds increased most rapidly when the response probability approached $g$ starting from a value of around 0.625. However, after reaching a value of around 0.5, the $D_T$ value for both groups with an SB mode of 60 seconds quickly overcame it and produced increasingly larger $D_T$ values as the response probability approached $g$. In terms of $D_{\mathrm{MR}}$, however, plot B shows that the group with an RRB percentage of 5\% and an SB mode of 22 seconds produced the greatest MR across most response probability values, regardless of the direction of association, and the group with an RRB percentage of 5\% and an SB mode of 60 seconds only marginally overcame it when the response probability was relatively close to $g$.

\subsection{Realised Level}

To investigate the usability of MaxMI-22 for estimating viable thresholds for RRB identification, we compared it to CUMP and MLN. We simulated large-scale assessment data at item level under the same conditions as the population-level inspection, except that the item difficulty parameter was limited to attain one of five values for simplicity: $b \in \{-2, -1, 0, 1, 2\}$. The sample size was set to either 500 or 2,500. The former is relatively small, whereas the latter is often encountered in large-scale assessments \citep[see][]{rios2024comparison}. Therefore, the total number of conditions was $2 \times 2 \times 2 \times 5 \times 2 = 80$. Each condition was repeated 100 times. The data generation procedure is described in Appendix~\ref{app_data_generation}.

In Appendix~\ref{app_bias_and_efficiency_of_MaxMI-22}, we present the bias and efficiency results of MaxMI-22. In short, we found that MaxMI-22 produced more inaccurate estimates of the thresholds when the conditional marginal probability of a correct response for SB test takers was closer to $g$. We also found that accuracy could be improved by smoothing the empirical MIRT curve using a unimodal monotone regression-based~\citep[e.g.,][]{frisen_unimodal_1986} approach prior to identifying the local maximum and estimating the threshold\footnote{This approach assumes unimodality of $\mathrm{P}_N(U_j=1\mid L_j=l_{ij})$ ($=u_{ij}$) as a function of the time threshold. Smoothed versions of $\mathrm{P}_N(U_j = 1 \mid X_{ij}=1)$ and $\mathrm{I}_N(X_{ij},U_j)$ can be constructed using the smoothed $\mathrm{P}_N(U_j=1 \mid L_j=l_{ij})$. The details are presented in Appendix~\ref{app_empirical_MIRT_smoothing_in_simulation}. OpenAI's GPT-5.2 model~\citep{openai_2025_gpt5.2} assisted in writing the R code for the pool-adjacent-violators algorithm~\citep[PAVA; see, for example,][]{barlow_1972_pava} used in unimodal monotone regression.}. Therefore, the results presented below are based on this approach.

First, we examined the proportions of the estimated thresholds across all conditions and repetitions. MaxMI-22 did not encounter any situations in which a threshold could not be estimated. In contrast, CUMP and MLN encountered several such situations, with 72.9\% and 77.7\% of thresholds being estimated across all conditions and repetitions, respectively. For CUMP, these situations occurred when the CUMP curve never crossed the chance level line. For MLN, however, such situations were cases where a two-component model did not fit the data better than a one-component model (1.2\%), where the estimated parameters did not align with the assumptions about the conceptual model underlying the RTs when RRB is present (5\%), or where the estimated RT density did not exhibit bimodality (16.1\%). Appendix~\ref{app_effects_of_parameters_on_CUMP_and_MLN} contains a figure depicting the effects of the population characteristics on the proportions of estimated thresholds.

\begin{figure}[!t]
\centering
\includegraphics[width=0.8\textwidth]{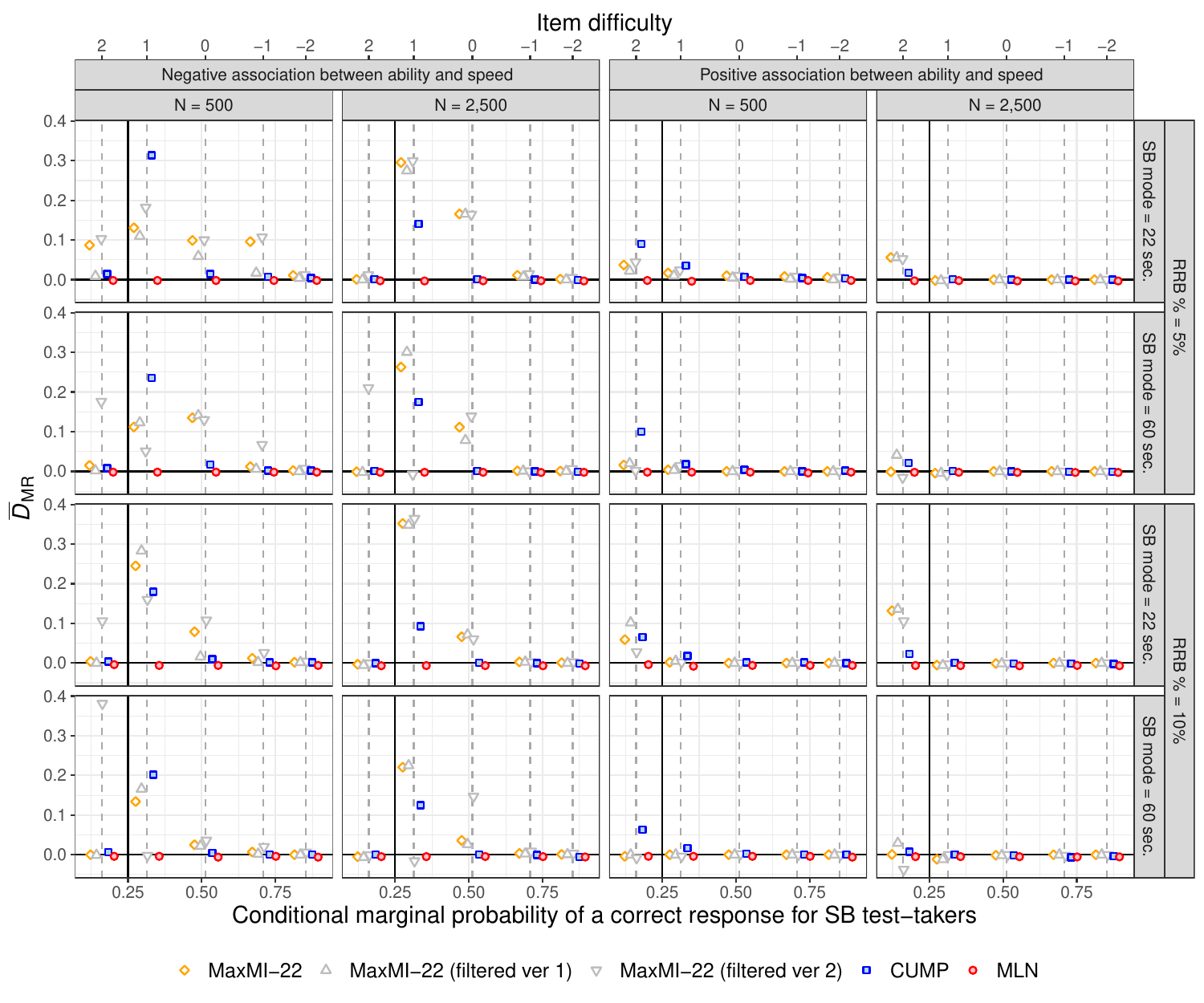}
\caption{Medians of the differences between the misclassification rates (MRs) produced by the MaxMI-22, CUMP, and MLN methods and the minimum MRs in the generated samples ($\overline{D}_{\mathrm{MR}}$) as functions of the population-level conditional marginal probability of a correct response for SB test takers. Note: N = sample size; RRB \% = RRB percentage; SB mode = RT mode for SB test takers. Comparisons between the methods are biased due to the different numbers of estimated thresholds (MaxMI-22, CUMP, and MLN). Therefore, the filtered results for MaxMI-22 indicate $\overline{D}_{\mathrm{MR}}$ values computed from only the cases for which both CUMP and MLN estimated a threshold (ver 1) and only the cases for which either CUMP or MLN was not able to estimate a threshold (ver 2). A small amount of random noise was added to the horizontal position of the data points for visual clarity. The vertical, dashed lines roughly indicate the true locations of the data points for each unique value of the SB response probability}
\label{fig5}
\end{figure}

Second, we examined the medians of the differences between the MRs produced by the MaxMI-22, CUMP, and MLN methods and the minimum MRs in the generated samples\footnote{The MRs were based on the posterior probabilities of belonging to the engaged component ($\Delta_j = 1$) given the observed responses ($u_{ij})$ and log RTs ($l_{ij})$, that is, 
\begin{align*}
    \mathrm{P}_i(\Delta_j = 1 \mid u_{ij}, l_{ij}) = \frac{\mathrm{P}(U_j = u_{ij}, L_j = l_{ij},\Delta_j = 1)}{\mathrm{P}(U_j = u_{ij}, L_j = l_{ij})}, \, i\in\{1, ..., N\}.
\end{align*}
OpenAI's GPT-5.2 model~\citep{openai_2025_gpt5.2} assisted in writing R code for efficient computation of the posterior probability of being in the engaged state given the observed data.} 
($\overline{D}_{\mathrm{MR}}$) as functions of the population-level conditional marginal probability of a correct response for SB test takers (or item difficulty) in groups categorised by the direction of association between ability and speed for SB test takers, sample size, the percentage of RRB, and the RT mode for SB test takers (Figure~\ref{fig5}). The reader is advised to be cautious when making comparisons between the methods based on Figure~\ref{fig5}, given that the number of estimated thresholds differs between methods, as discussed earlier. Therefore, because the CUMP and MLN methods produced substantial numbers of missing values, we also plotted the median differences for MaxMI-22 based on only the cases for which both CUMP and MLN estimated a threshold (59.5\% of all cases) as well as only the cases for which either CUMP or MLN was not able to estimate a threshold (40.5\%). In Figure~\ref{fig5}, these results are denoted as 'MaxMI-22 (filtered ver 1)' and 'MaxMI-22 (filtered ver 2)', respectively.

First, as with the population-level results, the effect of the response probability on $\overline{D}_{\mathrm{MR}}$ produced by MaxMI-22 was evident when the direction of association between ability and speed was negative. In this case, $\overline{D}_{\mathrm{MR}}$ diverged from zero as the response probability approached $g$ from the right. However, when the response probability differed significantly from $g$ (i.e., the item difficulty equalled $-1$ or $-2$), MaxMI-22 produced thresholds that were very close to those that minimised MR. Additionally, when the direction of association between ability and speed was positive, only those cases in which the response probability was smaller than $g$ were problematic. Second, the MLN method produced thresholds close to the times points where the MR is minimised, which was expected based on the study of~\citep{holopainen2026}. Third, sample size did not appear to impact the $\overline{D}_{\mathrm{MR}}$ values significantly, except when the response probability was close to the chance level. In these cases, if $\rho_{\theta\tau} < 0$, the $\overline{D}_{\mathrm{MR}}$ produced by MaxMI-22 was larger when the sample size was larger. Conversely, the $\overline{D}_{\mathrm{MR}}$ produced by CUMP was smaller when the sample size was larger, regardless of the direction of association.

Finally, when only the cases for which both CUMP and MLN estimated a threshold were used, the $\overline{D}_{\mathrm{MR}}$ values of MaxMI-22 were mostly similar to the $\overline{D}_{\mathrm{MR}}$ values based on all cases. Furthermore, when only the cases for which either CUMP or MLN was not able to estimate a threshold were used, the $\overline{D}_{\mathrm{MR}}$ values were also mostly similar to the $\overline{D}_{\mathrm{MR}}$ values based on all cases (with few exceptions; e.g., when $\rho_{\theta\tau} < 0$ and the SB response probability was below the chance level, i.e., the item difficulty equalled $2$). These results suggest that the problematic cases for CUMP and MLN did not affect the results of MaxMI-22 substantially. This is an encouraging result, because it indicates that the MaxMI-22 method may overcome the empirical limitations of the CUMP and MLN methods, provided that the SB response probability notably differs from that of the RRB test takers, where the required magnitude of the difference between the response probabilities depends on the direction of association between ability and speed.

\subsection{Summary}

The results suggested that, if the aim of identifying RRB is to minimise the misclassification rate, MaxMI-22 is beneficial compared to the CUMP and MLN methods when the conditional marginal probability of a correct response for SB test takers differs significantly from that for RRB test takers (i.e., the chance level, $g$). However, if these probabilities are similar, MaxMI-22 may produce a high misclassification rate. In these cases, some other characteristics of the population or data may affect the MIRT curve so that the maximum point does not distinguish between RRB and SB groups, but between some other groups. In other words, the greatest information gain does not come from the two explicit components, but from sub-components within the RRB and SB components, for example. Nevertheless, an advantage of MaxMI-22 was that it managed to produce thresholds in all cases, whereas CUMP and MLN did not produce thresholds in many cases. Conversely, if the aim is to identify time thresholds at which the data optimally splits into groups from an information-gain perspective, methods within the MaxMI framework are recommended. However, in this case, the resulting thresholds may not be suitable for identifying RRB specifically.

\section{Discussion}

In this paper, we proposed a non-parametric, \textit{mutual information} (MI)-based framework for identifying \textit{rapid-responding behaviour} (RRB) in computer-based educational and psychological measurements: the \textit{Maximal Mutual Information} (MaxMI) framework. The methods in this framework differentiate RRB from \textit{solution behaviour} (SB) by maximising the MI of the response and the (discretised) RT (MIRT) within a predefined search space as a function of time thresholds. The only differences between the methods are the number of categories in the relevant variables and the number of thresholds. The response variable can be treated as either the actual raw response (possibly nominal with more than two categories) or the response’s correctness (binary). The RT variable can be discretised into any reasonable number of categories with a varying number of thresholds.

We introduced three methods explicitly within the MaxMI framework: The first method uses response correctness and binarised RT and is abbreviated as 'MaxMI-22'; the second method uses correctness and categorises RT into three groups; and the third method uses raw response and binarised RT. We applied and analysed all the three methods using the PISA 2022 mathematics data set. Additionally, we investigated the behaviour and usability of MaxMI-22 in certain realistic conditions according to the \textit{Independent Latent Class Item Response Theory} (ILC-IRT) framework for response engagement of~\citet{nagy2022multilevel}, both at the population and realised levels.

In short, the empirical results suggested that the MaxMI framework is a viable alternative to existing methods for identifying RRB, and is often more suitable for this purpose, for at least three reasons: First, a MaxMI method will always identify a threshold as long as both the response and RT variable are non-deterministic. Second, the MaxMI framework can be used even if response correctness is not available or does not contain enough information to differentiate between RRB and SB or the observed RT distribution is not bimodal. Third, the framework may overcome the empirical limitations of the CUMP and MLN methods when they are unable to compute thresholds due to certain characteristics in the population or in the observed sample. However, if the distribution is a mixture of RRB and SB but the conditional marginal probabilities of a correct response for RRB and SB test takers are too similar, the MaxMI-22 method may produce an unsuitable threshold for differentiating between the two components. In addition to the empirical results, other advantages can be identified based on the definition of the framework. First, the framework is non-parametric. Second, the framework could be generalised to identify sub-groups other than RRB in achievement tests, as well as to identify disengaged responses in educational and psychological tests that do not measure achievement, such as questionnaires. Finally, the computed thresholds are optimal from an information-gain perspective.

This line of research still has much to offer. First, the behaviour of MaxMI methods other than MaxMI-22 at the population level remains to be seen. The following questions remain unanswered: When the population is characterised by ILC-IRT with two explicit components in the distribution, could one of the MaxMI-32 thresholds be suitable for identifying RRB? Could one of the thresholds also be used to split RRB or SB test takers into two sub-groups? In addition, how could a population of raw responses be modelled to study MaxMI-2Q? \citet{bock_1972_nrm}'s nominal response model could be useful for this purpose. If a suitable model is discovered, how would MaxMI-2Q behave? Finally, could other methods within the MaxMI framework, such as MaxMI-3Q or MaxMI-42, be viable alternatives?

Second, despite the other MaxMI methods, a better understanding of the behaviour of MaxMI-22 than was achieved in this study would be beneficial. This may be difficult to achieve analytically because MIRT is a complicated function of the time threshold. However, analysing MIRT as a function of the pair $\{\mathrm{P}(X=1,Y=1), \mathrm{P}(X=1)\mathrm{P(Y=1)}\}$ could be useful because we know that MI is convex in this pair~\citep[][Theorem 2.7.2]{cover1991elements}. Additionally, we know that MI is a convex function of $\mathrm{P}(X=x \mid Y=y)$ for a fixed $\mathrm{P}(Y=y)$~\citep[][Theorem 2.7.4]{cover1991elements}. In our case, $\mathrm{P(Y=y)}$ is indeed fixed as a function of the time threshold.

Future research could also improve our understanding simply by modifying the ILC-IRT model used in this study to investigate the behaviour of MaxMI-22. For example, other types of sub-groups could be implemented as explicit components in the distribution~\citep[e.g., rapid versus semi-rapid responding, or partial engagement; see, for example,][]{wise2017rgb,wise_kuhfeld_2021}. Other models could also be considered for RT, such as the Weibull model~\citep{rouder_2003}, and the joint modelling of responses and RTs could be approached differently, for example using semi-parametric models~\citep[e.g.,][]{liu_2025,wang_semiparam_2013}, race models~\citep{rouder_2015}, or bivariate generalised IRT models~\citep{molenaar_2015}.

\section*{Acknowledgments}

The authors designed this study, conducted the analyses, interpreted the results, and wrote the article. OpenAI's GPT-5.2 model was used to suggest some references. GPT-5.2 also assisted in writing the following R code: 1) user-defined functions for conducting unimodal monotone regression (19 May 2026) and 2) user-defined functions for efficient computation of the posterior probability of being in the engaged state given the observed data (23 April 2026). DeepL's writing assistant was used to improve the spelling, grammar, and flow of the article (25 September 2026). However, the authors reviewed all output generated by AI and take full responsibility for the integrity of the whole content.

\section*{Data Availability}

The PISA 2022 dataset is publicly available at the OECD website: \\ https://www.oecd.org/en/data/datasets/pisa-2022-database.html. The datasets generated during the current study can be re-generated with the provided R codes. The R codes that can be used to reproduce all empirical results are available on GitHub: https://github.com/sajomaho-uni/Empirical-Results-of-MaxMI-for-RRB-Identification.

\section*{Funding}

The present study is part of the EDUCA Flagship funded by the Research Council of Finland under Grants \#358924 and \#358947 and the EDUCA-Doc Doctoral Education pilot funded by the Ministry of Education and Culture under Grant \#VN/3137/2024-OKM-4 (Doctoral school pilot).

\begin{appendices}

\section{Appendix. The ILC-IRT Framework for Response Engagement} \label{app_ILC-IRT}

This Appendix introduces the the \textit{Independent Latent Class Item Response Theory} (ILC-IRT) framework for response engagement of~\citep{nagy2022multilevel}. In the ILC-IRT framework, RTs are viewed as indicators of engagement, and the distributions of responses and RTs are mixtures of engaged and disengaged distributions. The framework assumes that (1) responses on an item are conditionally independent given a person’s ability and behaviour type (e.g., RRB or SB), (2) RTs on an item are conditionally independent given a person’s speed and behaviour type, and (3) responses and RTs on an item are conditionally independent given the person parameters and behaviour type.

Let $\Delta_{ij}$ with support $\mathcal{D} = \{0,1\}$ be a random indicator variable for the response engagement of person $i$ on item $j$, with $\Delta_{ij} = 1$ indicating an engaged response and $\Delta_{ij} = 0$ a disengaged response. The probability of a person being engaged with an item has been parametrised in different ways: \citet{ulitzsch2020hierarchical} assumed a Rasch model, meaning that the probability of person $i$ being engaged with item $j$ is
\begin{equation} \label{eq_engagement_prob}
    \mathrm{P}(\Delta_{ij}=1\mid\phi_i, \iota_j)=\frac{\exp(\phi_i-\iota_j)}{1+\exp(\phi_i-\iota_j)},
\end{equation}
where $\phi_i$ is the latent engagement of person $i$ and $\iota_j$ is the engagement difficulty of item $j$. In contrast, \citet{wang2015mixture} and \citet{wang2018detecting} did not include person parameters for $\mathrm{P}(\Delta_{ij}=1)$; \citet{wang2015mixture} utilised one fixed effect and \citet{wang2018detecting} item effects. Going forward, we assume that the probability of being in the engaged state is parametrised according to the Rasch model in Eq.~\eqref{eq_engagement_prob}.

Let $U_{ij}$ with support $\mathcal{U} = \{0,1\}$ denote the random variable for response correctness of person $i$ on item $j$. In the ILC-IRT framework, it is assumed that disengaged responses are not related to ability and their accuracy corresponds to the chance level, that is, 
\begin{equation} \label{eq_cor_resp_prob_disengaged_comp}
    \mathrm{P}(U_{ij} = 1\mid\Delta_{ij} = 0, g_j) = g_j,
\end{equation}
where $g_j$ is the chance level for item $j$. In contrast, the probability of a correct engaged response is parametrised according to an IRT model with person and item parameters: 
\begin{equation} \label{eq_cor_resp_prob_engaged_comp}
    \mathrm{P}(U_{ij} = 1\mid\Delta_{ij} = 1, \theta_i, \boldsymbol{\xi}_j) = p(\theta_i, \boldsymbol{\xi}_j),
\end{equation}
where $\theta_i$ is the latent ability of person $i$ and $\boldsymbol{\xi}_j$ is the vector of item parameters (e.g., difficulty, discrimination, pseudo-guessing). In the general IRT framework for response engagement, different models have been utilised, such as the one-parameter logistic IRT model \citep[1PL;][]{ulitzsch2020hierarchical}, the two-parameter model \citep[2PL;][]{pokropek2016grade}, and the three-parameter model \citep[3PL;][]{wang2015mixture,wang2018detecting}.

Let $L_{ij}$ with support $\mathcal{L} = \mathbb{R}$ denote the random variable for the log RT of person $i$ on item $j$. In the ILC-IRT framework, it is often assumed that both disengaged and engaged log RTs follow a normal distribution, but with different parametrisations. For example, \citet{schnipke1997modeling}, \citet{ulitzsch2020hierarchical}, and \citet{wang2015mixture} assumed that the mean and variance of the disengaged log RTs were common across individuals and items: 
\begin{equation} \label{eq_log_RT_dist_disengaged}
    (L_{ij} \mid \Delta_{ij}=0, \mu_c, \sigma_c) \sim N(\mu_c, \sigma_c^2),
\end{equation}
where $\mu_c$ and $\sigma_c^2$ are the mean and variance, respectively. In contrast, the mean and variance of the engaged log RTs may contain person and item parameters. For example, 
\begin{equation} \label{eq_log_RT_dist_engaged}
    (L_{ij} \mid \Delta_{ij}=1, \tau_i, \alpha_j, \beta_j) \sim N(\beta_j - \tau_i, \alpha_j^2),
\end{equation}
where $\tau_i$ is the latent speed of person $i$, and $\beta_j$ and $\alpha_j$ are the time intensity and discrimination parameters of item $j$, respectively \citep[e.g.,][]{ulitzsch2020hierarchical,wang2015mixture,wang2018detecting}.

To facilitate the computation of MIRT in the ILC-IRT framework, we define an indicator variable for the log RT of person $i$ on item $j$ as a function of $L_{ij}$ with
\begin{equation} \label{eq_x_def}
X_{ij} = \mathrm{I}_{\mathcal{L}}(L_{ij}) = 
    \begin{cases}
    1\mathrm{, if }\;L_{ij} \leq l_0, \\
    0\mathrm{, otherwise,}
    \end{cases}
\end{equation}
where $l_0 \in \mathcal{L}$ is a fixed time point. Consequently, using the parametrisations for log RT in \eqref{eq_log_RT_dist_disengaged} and \eqref{eq_log_RT_dist_engaged}, the probabilities of person $i$ having a log RT of less than or equal to $l_0$ to item $j$ while being in the disengaged state, as well as in the engaged state, are
\begin{align} \label{eq_x_cond_probs}
    \mathrm{P}(X_{ij}=1\mid\Delta_{ij}=0,\mu_c,\sigma_c) &= \Phi\left(\frac{l_0 - \mu_c} {\sigma_c}\right) \ \mathrm{and} \\
    \mathrm{P}(X_{ij}=1\mid\Delta_{ij}=1,\tau_i,\beta_j,\alpha_j) &= \Phi\left(\frac{l_0 - \beta_j + \tau_i} {\alpha_j}\right),
\end{align}
respectively, where $\Phi$ denotes the cumulative density function of a standard normal distribution.

\begin{figure}[t]
\includegraphics[width=0.9\textwidth]{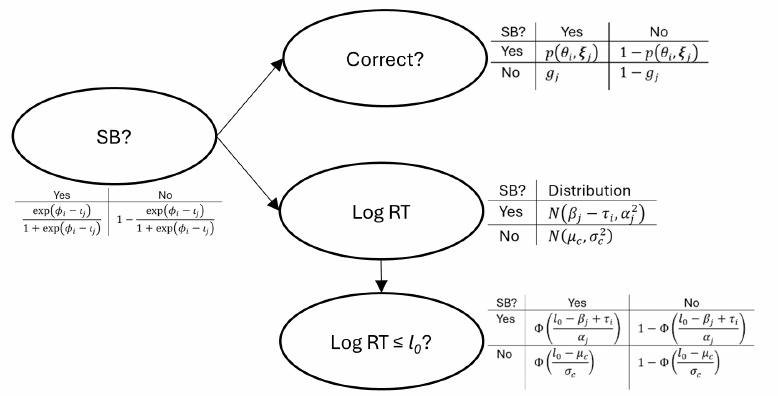}
\caption{Illustration of the ILC-IRT framework}
\label{fig6}
\end{figure}

The person parameters are assumed to follow a multivariate normal distribution with a mean vector
\begin{equation} \label{eq_person_par_mean_vec}
    \boldsymbol{\mu}_{\mathcal{P}} =
    \begin{pmatrix}
    \mu_\phi & \mu_\theta & \mu_\tau
    \end{pmatrix}
\end{equation}
and covariance matrix
\begin{equation} \label{eq_person_par_cov_mat}
\boldsymbol{\Sigma}_{\mathcal{P}} = 
    \begin{pmatrix}
    \sigma_{\phi}^2 & \sigma_{\phi\theta} & \sigma_{\phi\tau} \\
    \sigma_{\phi\theta} & \sigma_{\theta}^2 & \sigma_{\theta\tau} \\
    \sigma_{\phi\tau} & \sigma_{\theta\tau} & \sigma_{\tau}^2
    \end{pmatrix}.
\end{equation}
Note that the distribution is reduced to a bivariate case if $\mathrm{P}(\Delta_{ij} = 1)$ is not affected by the person. Figure~\ref{fig6} illustrates the framework.

\section{Appendix. Parameter Selections for the Investigations of MaxMI in the ILC-IRT Framework} \label{app_parameter_selections}

In this Appendix, we describe the selection of the parameters for the population level examinations as well as the simulation study. The means of the person parameters were set to zero ($\mu_\phi=0$, $\mu_\theta = 0$, and $\mu_\tau=0$). Following the selections of~\citet{ulitzsch2020hierarchical} in their simulation study, the variances of the person parameters were fixed to $\sigma_{\phi}^2 = 3.5$, $\sigma_{\theta}^2 = 1$, and $\sigma_{\tau}^2 = 0.05$ and the correlations (covariances) of engagement with ability and speed to $\rho_{\phi\theta} = 0.55~(\sigma_{\phi\theta} \approx 1.03)$ and $\rho_{\phi\tau} = 0.20~(\sigma_{\phi\tau} \approx 0.08)$, respectively. The correlation (covariance) of ability with speed was set to either $\rho_{\theta\tau} = -0.40~(\sigma_{\theta\tau} \approx -0.09)$ or $\rho_{\theta\tau} = 0.40~(\sigma_{\theta\tau} \approx 0.09)$, of which the former was consistent with~\citet{ulitzsch2020hierarchical}, but the latter was added as an additional condition to reflect situations in large-scale assessments where the correlation between ability and speed is positive~\citep[see, for example,][]{holopainen_gauging_2025,holopainen2026}.

The engagement difficulty parameter, $\iota_j$, was allowed to attain values of $-4.27$ and $-3.29$, corresponding to approximately 5\% and 10\% RRB, respectively, with the given distribution of the person parameters. The smallest $\iota_j$ value represents items that are very easy to be engaged with, e.g., items at the beginning of a test or items that do not require much reading. The largest value represents the average RRB proportion of 10\% in low-stakes assessments found by~\citet{rios2022meta}.

In accordance with~\citet{ulitzsch2020hierarchical}, we used the unidimensional 1PL response model for engaged responses, which meant that the probability of a correct engaged response was
\begin{equation}
    p(\theta_i, b_j) = \frac{\exp(\theta_i-b_j)}{1+\exp(\theta_i-b_j)},
\end{equation}
where $b_j$ represents the difficulty parameter of item $j$. In the population-level examinations, $b$ was allowed to attain values of $-4 + \frac{8k}{99}$, $k \in \{0, ..., 99\}$. In the realised-level examinations, $b$ was allowed to attain values of $-2 + k$, $k \in \{0, ..., 4\}$. In contrast, the probability of a correct disengaged response was $g_j = 0.25$.

Following the selections of~\citet{ulitzsch2020hierarchical} in their simulation study, the time intensity parameter, $\beta_j$, was allowed to attain values of 3.25 and 4.25, but the time discrimination parameter was set to $\alpha_j = \sqrt{0.15}$ for all items. Finally, the mean and variance of the disengaged log RTs were set to $\mu_c = 3$ and $\sigma_c^2 = 1.95$. All parameters are gathered in Table~\ref{table4}.

\begin{table}[ht]
\begin{center}
\caption{Parameters for the Investigations of MaxMI-22 in the ILC-IRT Framework\label{table4}}
\begin{tabular}{@{}lll@{}}
\toprule
Parameter & \multicolumn{1}{c}{Parameter explanation} & \multicolumn{1}{c}{Value(s)} \\
\midrule
\multicolumn{3}{c}{Person parameters} \\
$\mu_\phi$ & Mean of the latent engagement tendency & 0 \\
$\sigma_\phi^2$ & Variance of the latent engagement tendency & 3.5 \\
$\mu_\theta$ & Mean of the latent ability & 0 \\
$\sigma_\theta^2$ & Variance of the latent ability & 1 \\
$\mu_\tau$ & Mean of the latent speed & 0 \\
$\sigma_\tau^2$ & Variance of the latent speed & 0.05 \\
$\rho_{\phi\theta}$ & Correlation between engagement and ability & 0.55 \\
$\rho_{\phi\tau}$ & Correlation between engagement and speed & 0.20 \\
$\rho_{\theta\tau}$& Correlation between ability and speed & $-0.40$, 0.4 \\
\multicolumn{3}{c}{Item parameters} \\
$\iota_j$ & Engagement difficulty & $-4.27$, $-3.29$ \\
$g_j$ & Chance level & 0.25 \\
$b_j$ & Item difficulty & $-4 + \frac{8k}{99}$, $k \in \{0, ..., 99\}$\textsuperscript{a} or \\
& & \quad $-2 + k$, $k \in \{0, ..., 4\}$\textsuperscript{b} \\
$\mu_c$ & Mean of the disengaged log RTs & 3 \\
$\sigma_c^2$ & Variance of the disengaged log RTs & 1.95 \\
$\beta_j$ & Time intensity & 3.25, 4.25 \\
$\alpha_j$ & Time discrimination & $\sqrt{0.15}$ \\
\bottomrule
\end{tabular}
\begin{tablenotes}
\item Note: \textsuperscript{a} For population-level investigations.
\item \textsuperscript{b} For realised level investigations.
\end{tablenotes}
\end{center}
\end{table}

\section{Appendix. Data Generation in the Simulation Study} \label{app_data_generation}

Data generation was carried out in three steps. In the first step, we generated the person parameters from a multivariate normal distribution using the 'MASS' package in R~\citep{venables2002mass_package}. In the second step, we generated the engagement indicator for each person: Person $i$ was engaged on item $j$ if a random number drawn from $U(0,1)$ was less than or equal to the engagement probability, $P(\Delta_{ij}=1 \mid \phi_i,\iota_j)$ (1PL IRT), and disengaged otherwise. In the third step, we generated the responses and log RTs for each person: If person $i$ was engaged on item $j$, their response was correct if a random number drawn from $U(0,1)$ was less than or equal to $p(\theta_i, b_j)$ (1PL IRT), and incorrect otherwise, and their log RT was generated from $N(\beta_j -\tau_i, \alpha_j^2)$. However, if person $i$ was disengaged on item $j$, their response was correct if a random number drawn from $U(0,1)$ was less than or equal to $g_j$, and incorrect otherwise, and their log RT was drawn from $N(\mu_c,\sigma_c^2)$.

\section{Appendix. CUMP and MLN Thresholds in the PISA 2022 Example and the Simulation Study} \label{app_CUMP_and_MLN_in_PISA_and_simulation}

This Appendix describes the operational rules used to apply the CUMP~\citep{guo2016cump} and MLN~\citep{rios2020mln} methods to the PISA 2022 data as well as the generated data in the simulation study. The CUMP method assumes that rapid responses are random. If this is the case, the conditional proportion of correct responses should vary around the chance level at small RTs. That is, CUMP is a parametric method in the sense that RRB accuracy is modelled via an item-specific parameter. While previous research have demonstrated that rapid responses are random~\citep[e.g.,][]{guo2016cump,lee2014vitp}, more recent studies have also shown that it may not always be the case~\citep[e.g.,][]{holopainen_gauging_2025,holopainen2026,wise_kuhfeld_2020}. 

\citet{holopainen2026} provided a modified formula for calculating the CUMP thresholds, which considers the possibility that some items are so difficult that the overall proportion of correct responses is less than the chance level. The formula is
\begin{align} \label{eq_cump_threshold_formula}
    C_j = \begin{cases}
    \mathrm{max}\{t_{ij}, \, i \in \{1, ..., N\}{:}\;\mathrm{CUMP}_j(t_{ij})\leq g_j\},\;\mathrm{if}\;P^{+}_j>g_j, \\
    \mathrm{max}\{t_{ij}, \, i \in \{1, ..., N\}{:}\;\mathrm{CUMP}_j(t_{ij})\geq g_j\},\;\mathrm{if}\;P^{+}_j<g_j, \\
    \mathrm{not\;defined,\;if}\; P^{+}_j=g_j,
    \end{cases}
\end{align}
where $i$ indexes the observations, $N$ is the sample size, $P^{+}_j$ is the overall proportion of correct responses of item $j$ in the sample, $\mathrm{CUMP}_j(t_{ij})$ is the CUMP curve, i.e., the proportion of correct responses of all those test takers who spent $t_{ij}$ time units or fewer on item $j$, and $g_j$ is the chance level of item $j$. Note that in this case $P^{+}_j = \mathrm{P}_N(U_j = 1)$ and $\mathrm{CUMP}_j(t_{ij}) = \mathrm{P}_N(U_j = 1 \mid T_{j} \leq t_{ij})$. Because all considered items were multiple-choice items with four response options, the chance level in the CUMP formula was set naturally to $g = 0.25$ for all cases. In each case, first ten observations were omitted to reduce noise in the CUMP curve before using the formula in Eq.~\eqref{eq_cump_threshold_formula}.

The MLN method first fits a two-component mixture model to the log RT data, based on which the estimated RT density is
\begin{equation}
    \hat{p}(t) = \left[\hat{\lambda}_1f_1\left(\log(t);\hat{\mu}_1,\hat{\sigma}_1^2\right) + \left(1-\hat{\lambda}_1\right)f_2\left(\log(t);\hat{\mu}_2,\hat{\sigma}_2^2\right)\right] / t,
\end{equation}
where $t$ is the realised value of RT, $\hat{\lambda}_1$ is the estimated mixture weight of the first component, $\hat{\mu}_1$, $\hat{\sigma}_1$, $\hat{\mu}_2$, and $\hat{\sigma}_2$ are the estimated means and standard deviations, and $f_1$ and $f_2$ are normal distribution functions. The MLN threshold is defined as the time point between the two RT modes, $\exp{(\hat{\mu}_1-\hat{\sigma}_1^2)}$ and $\exp{(\hat{\mu}_2-\hat{\sigma}_2^2)}$, where $\hat{p}(t)$ reaches its minimum value.

However, before fitting the model to the data and estimating an MLN threshold, there were three issues to be considered: First, in each case, we needed to assess the number of components in the observed RT distribution. Following~\citet{holopainen2026}, we did this by comparing the fits of a one- and a two-component mixture model to the log RT data assuming normal distributions based on the Bayesian Information Criterion~\citep[BIC;][]{schwarz1978}; if the two-component model fit the data better, the distribution was deemed bimodal and we proceeded with the threshold estimation, whereas if the one-component fit the data better, or the two-component model's estimation did not converge to a solution, the distribution was deemed unimodal and we did not proceed. We used the R package "mixtools"~\citep{benaglia2009} to fit the two-component models, where the algorithm was allowed to randomly generate the initial mean values. Second, the estimated parameters of the two-component model needed to align with the assumptions about the conceptual model underlying the RTs when RRB is present: RRB proportion needed to be less than 0.5 and the RT mode in the RRB component needed to be less than the RT mode in the SB component (we determined the RRB component as the component with the smaller log RT mode). If the preceding conditions applied, we proceeded with the threshold estimation, but otherwise, we did not proceed. Third, because MLN assumes that a local minimum of the estimated RT density can be found on an open interval between the RT modes, the method is not justified in cases where one of the modes remains hidden with respect to the estimated RT density, even if such a two-component mixture model solution could be found that fit the data better than a one-component solution. Therefore, in these cases, the MLN method could not estimate a viable threshold. See, for example, \citet{holopainen2026}, who also encountered this issue in their simulation study.

\section{Appendix. Smoothing of the Empirical MIRT in the Simulation Study} \label{app_empirical_MIRT_smoothing_in_simulation}

This Appendix describes the smoothing procedure applied to the empirical MIRT before using the MaxMI-22 method on the generated data in the simulation study. Prior analysis revealed that the empirical MIRT is very noisy in some situations, resulting in inaccurate estimates of the population-level threshold value (see Appendix~\ref{app_bias_and_efficiency_of_MaxMI-22} for the bias and efficiency results). In these cases, the empirical MIRT curve can be smoothed before finding the global maximum in the predefined search space. For instance, if we assume unimodality of $\mathrm{P}_N(U_j=1\mid L_j=l_{ij})$ ($=u_{ij}$) as a function of the time threshold, the curve can be smoothed with unimodal monotone regression\footnote{Note that the resulting "smoothed" curve is not a smooth curve per se, but rather a stepwise function of the time threshold.} \citep[UMR; e.g.,][]{frisen_unimodal_1986}, and consequently, smoothed versions of $\mathrm{P}_N(U_j = 1 \mid X_{ij}=1)$ and $\mathrm{I}_N(X_{ij},U_j)$ can be constructed using the smoothed $\mathrm{P}_N(U_j=1 \mid L_j=l_{ij})$. We performed UMR using our own implementation that utilises the pool-adjacent-violators algorithm~\citep[PAVA; e.g.,][]{barlow_1972_pava}. Our approach follows that of \citet{busing2022monotone}, but rather than finding the location of the mode by minimizing the squared error, our method locates the mode by maximizing likelihood (ML). In addition, the mode may be an actual mode or an antimode. The selection between these two fits was also done based on ML: UMR with a mode (antimode) was the best fit to the data for 89\% (11\%) of all cases.

\section{Appendix. Investigation of the Bias and Efficiency of the MaxMI-22 Method} \label{app_bias_and_efficiency_of_MaxMI-22}

\begin{figure}[!t]
\centering
\includegraphics[width=0.8\textwidth]{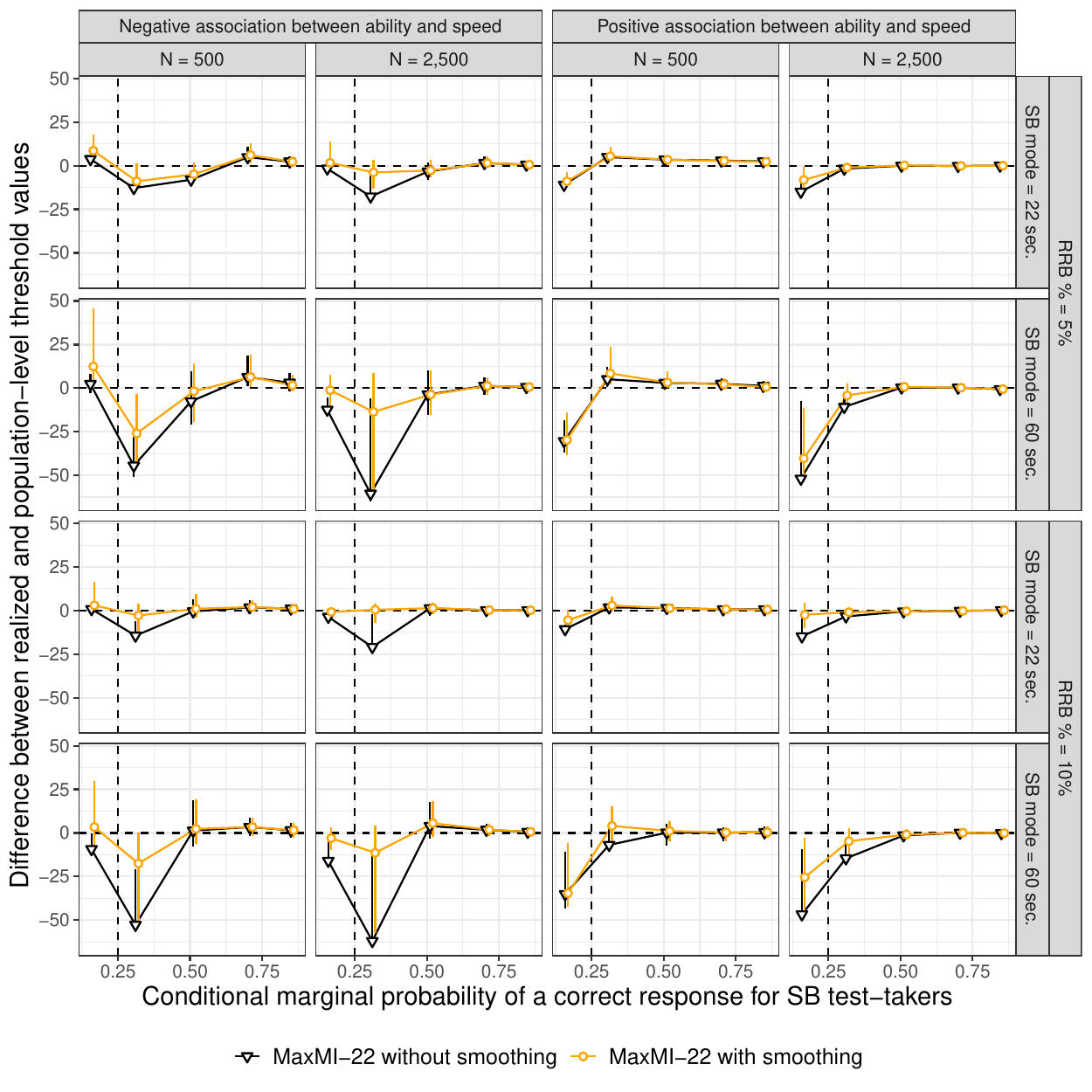}
\caption{Medians and interquartile ranges of the differences between the time thresholds estimated by MaxMI-22 and the true population-level threshold values (in seconds) as functions of the population-level conditional marginal probabilities of a correct response for SB test takers. Note: N = sample size; RRB \% = RRB percentage; SB mode = RT mode for SB test takers}
\label{fig7}
\end{figure}

Before the main analysis, we analysed the bias and efficiency of the MaxMI-22 method with and without smoothing the empirical MIRT curve before threshold estimation (see Appendix~\ref{app_empirical_MIRT_smoothing_in_simulation} for the explanation of the smoothing procedure). Figure~\ref{fig7} shows the medians and interquartile ranges of the differences between the estimated and population-level threshold values. We notice that MaxMI-22 without smoothing produced more inaccurate estimates of the thresholds the closer the conditional marginal probability of a correct response for SB test takers was to that of the RRB test takers (i.e., the chance level, $g=0.25$). However, smoothing the empirical MIRT before threshold estimation enhanced the accuracy of the thresholds in most of these problematic cases. However, when the association between the latent ability and speed was negative and the SB response probability was smaller than the chance level, the MaxMI-22 method with smoothing produced less accurate and efficient thresholds than the version without smoothing. Nevertheless, we opted to use MaxMI-22 with smoothing before threshold estimation in the main analysis.

\section{Appendix. Effects of the Population Characteristics on the Proportions of the Estimated CUMP and MLN Thresholds} \label{app_effects_of_parameters_on_CUMP_and_MLN}

\begin{figure}[!t]
\centering
\includegraphics[width=0.8\textwidth]{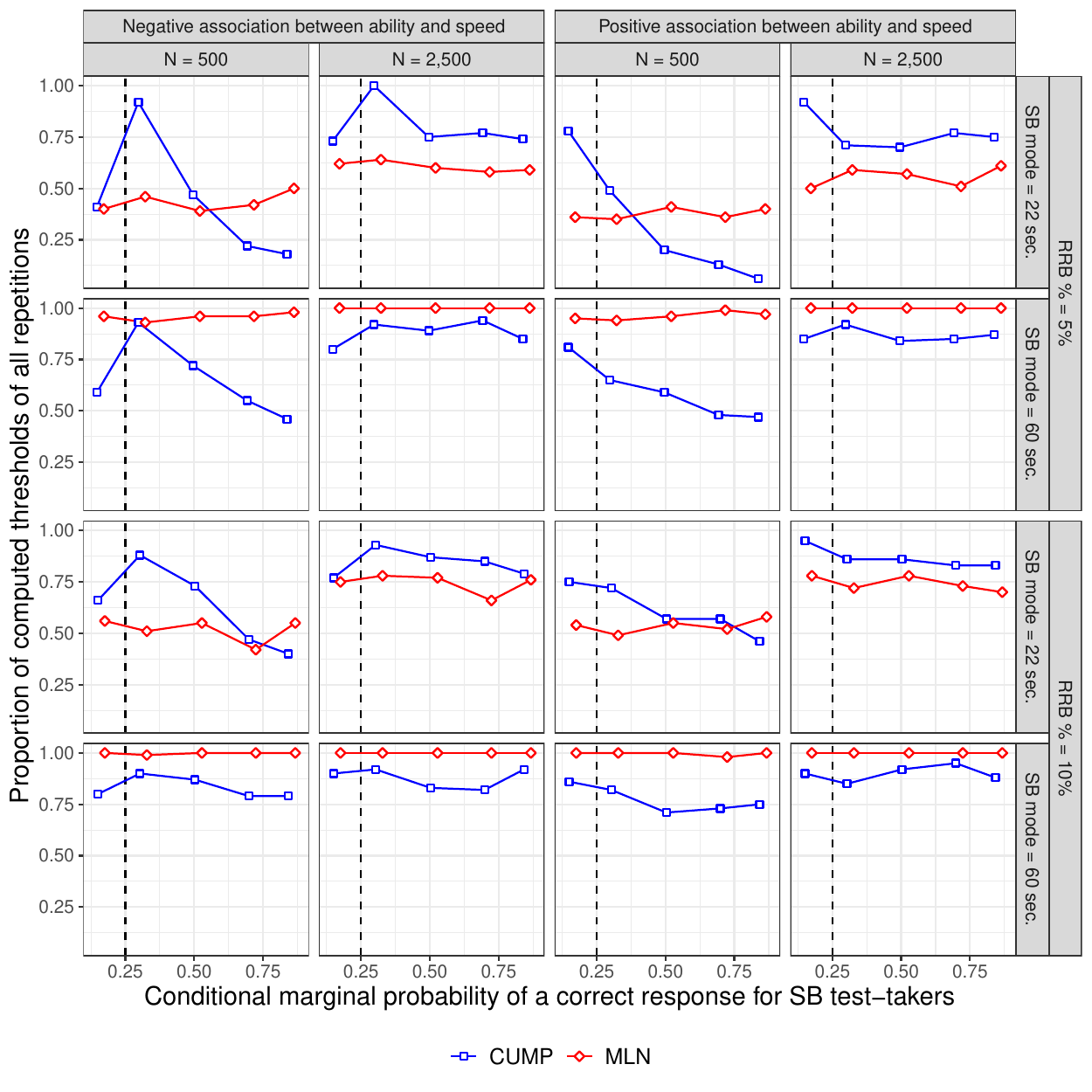}
\caption{Proportions of estimated thresholds of all repetitions as functions of the population-level conditional marginal probabilities of a correct response for SB test takers. Note: N = sample size; RRB \% = RRB percentage; SB mode = RT mode for SB test takers}
\label{fig8}
\end{figure}

Figure~\ref{fig8} illustrates the effects of the population characteristics on the proportions of estimated thresholds by CUMP and MLN. The ability to estimate thresholds of both methods were affected negatively by a smaller sample size. In addition, when sample size was smaller, CUMP was affected negatively by a larger conditional marginal probability of a correct response for SB test takers, as can be seen by the threshold proportion decreasing as the response probability increases in most cases. Meanwhile, regardless of sample size, MLN was affected negatively by a smaller SB mode.

\end{appendices}

\clearpage

\bibliographystyle{apalike}
\bibliography{references}

\newpage
\renewcommand{\notesname}{Notes}
\theendnotes

\end{document}


\maketitle

\newpage

\section{Description of This Supplementary Material}

This supplementary material is for the preprint titled, 'Mutual Information as a Tool for Optimal Classification: Application to Identifying Rapid-Responding Behaviour'. In general, the material appends the preprint by discussing the ILC-IRT framework further. First, in Section~\ref{sec_MIRT_in_ILC-IRT}, we implement MIRT and MaxMI into the ILC-IRT framework. Second, in Section~\ref{sec_CUMP_and_MLN_in_ILC-IRT}, we do the same for the CUMP and MLN methods and show that they are not theoretically justified in the ILC-IRT. Finally, in Section~\ref{sec_MR}, we show the formulation for the misclassification rate (MR) of the log RT indicator and discuss the potential shape of the MR as a function of the time threshold.

\section{MIRT and MaxMI in the ILC-IRT Framework} \label{sec_MIRT_in_ILC-IRT}

In this section, we bring MIRT into the ILC-IRT framework as a measure of non-linear dependency between the response and binarized RT variables. To achieve this, we must derive the marginal probabilities $\mathrm{P}(\Delta_j=1 \mid \iota_j)$, $\mathrm{P}(U_j=1 \mid \iota_j, \boldsymbol{\xi}_j,g_j)$, $\mathrm{P}(X_j=1 \mid \iota_j, \alpha_j, \beta_j, \mu_c, \sigma_c)$, and $\mathrm{P}(X_j=1,U_j=1 \mid \iota_j, \boldsymbol{\xi}_j, g_j, \alpha_j, \beta_j, \mu_c, \sigma_c)$ and the marginal density of log RT for a population of test-takers according to the ILC-IRT framework. We also show that, due to the effect of $\phi$, the log RT density of the engaged component is not proportional to a normal density. To achieve these aims, we remove the effects of the person parameters by marginalizing over their underlying distributions. We simplify the integrals such that they are composed of one or more one-dimensional integrals. For the probability of being in the engaged state, we assume the Rasch model; for the probability of answering correctly, we assume the item-specific chance level probability for disengaged responses and an IRT model (e.g., 1PL, 2PL, or 3PL) for engaged responses; for log RT, we assume the normal distribution with common mean and variance across individuals (and items) for disengaged responses and the normal distribution with person and item parameters for engaged responses; for the person parameters, we assume the multivariate normal distribution.

However, we begin by defining the conditional mean and variance of a person parameter given the other person parameters, which helps us later define the probability distributions. For instance, the conditional mean and variance of $\tau$ given $\phi$ and $\theta$ are
\begin{align} \label{eq_cond_mu_of_tau_given_phi_and_theta}
    \begin{aligned}
        \mu_{\tau\mid \phi,\theta} 
        = \mu_{\tau} +
        \begin{bmatrix}
        \sigma_{\phi\tau} & \sigma_{\theta\tau}
        \end{bmatrix}
        \begin{pmatrix}
        \sigma_{\phi}^2 & \sigma_{\phi\theta} \\
        \sigma_{\phi\theta} & \sigma_{\theta}^2
        \end{pmatrix}^{-1}
        \begin{bmatrix}
        \phi - \mu_{\phi}\\
        \theta - \mu_{\theta}
        \end{bmatrix}
    \end{aligned}
\end{align}
and
\begin{align} \label{eq_cond_sigma_squared_of_tau_given_phi_and_theta}
    \begin{aligned}
        \sigma_{\tau\mid \phi,\theta}^2 
        = \sigma_{\tau}^2 -
        \begin{bmatrix}
        \sigma_{\phi\tau} & \sigma_{\theta\tau}
        \end{bmatrix}
        \begin{pmatrix}
        \sigma_{\phi}^2 & \sigma_{\phi\theta} \\
        \sigma_{\phi\theta} & \sigma_{\theta}^2
        \end{pmatrix}^{-1}
        \begin{bmatrix}
        \sigma_{\phi\tau} \\
        \sigma_{\theta\tau}
        \end{bmatrix},
    \end{aligned}
\end{align}
respectively. The other conditional means and variances are defined similarly.

First, let $f_{\phi}$ denote the density function of $\phi$. The probability of a random test-taker from the population responding to item $j$ under the engaged state is
\begin{align} \label{eq_engagement_prob_population}
    \begin{aligned}
    \mathrm{P}(\Delta_{j}=1\mid \iota_j) &= \int_{\mathbb{R}} \mathrm{P}(\Delta_j = 1 \mid  \phi, \iota_j)f_{\phi}(\phi; \mu_{\phi}, \sigma_{\phi}^2)d\phi \\
    & = \int_{\mathbb{R}}\frac{\exp(\phi-\iota_j)}{1+\exp(\phi-\iota_j)}f_{\phi}(\phi; \mu_{\phi}, \sigma_{\phi}^2) d\phi,
    \end{aligned}
\end{align}
which is a logistic-normal integral and does not have a closed-form solution, but numerical methods can be utilised to compute the integral, or it can be approximated~\citep{pirjol2013logistic_normal_int}. In this study, we computed most integrals with numerical integration and utilised the "cubature" package in R~\citep{cubature_package}. Second, let $f_{\phi, \theta} = f_{\theta\mid \phi}f_{\phi}$ denote the joint density function of $\phi$ and $\theta$ and $f_{\theta\mid \phi}$ the conditional density function of $\theta$ given $\phi$. The probability of a random test-taker being in the engaged state and answering item $j$ correctly is
\begin{align} \label{eq_marg_prob_mass_dist_of_y_engaged_population}
    \begin{aligned}
    & \mathrm{P}(U_j = 1, \Delta_j = 1\mid \iota_j, \boldsymbol{\xi}_j) \\
    & = \iint_{\mathbb{R}} \mathrm{P}(U_j = 1, \Delta_j = 1 \mid  \phi, \theta, \iota_j, \boldsymbol{\xi}_j) f_{\phi,\theta}(\phi, \theta; \mu_{\phi,\theta}, \Sigma_{\phi,\theta}) d\theta d\phi \\
    & = \iint_{\mathbb{R}} \mathrm{P}(\Delta_j = 1 \mid  \phi, \iota_j)\mathrm{P}(U_j = 1 \mid  \Delta_j = 1, \theta, \boldsymbol{\xi}_j) \\
    & \quad \times f_{\theta\mid \phi}(\theta; \mu_{\theta\mid \phi}, \sigma_{\theta\mid \phi}^2) f_{\phi}(\phi; \mu_{\phi}, \sigma_{\phi}^2) d\theta d\phi \\
    & = \int_{\mathbb{R}} \mathrm{P}(\Delta_j = 1 \mid  \phi, \iota_j) f_{\phi}(\phi; \mu_{\phi}, \sigma_{\phi}^2)\int_{\mathbb{R}} p(\theta, \boldsymbol{\xi}_j) f_{\theta\mid \phi}(\theta; \mu_{\theta\mid \phi}, \sigma_{\theta\mid \phi}^2) d\theta d\phi.
    \end{aligned}
\end{align}
In contrast, the probability of a random test-taker being disengaged and answering item $j$ correctly does not depend on the ability parameter, and thus
\begin{align} \label{eq_marg_prob_mass_dist_of_y_disengaged_population}
    \mathrm{P}(U_j = 1, \Delta_j = 0 \mid \iota_j, g_j) = \mathrm{P}(\Delta_j = 0)g_j,
\end{align}
where $g_j$ is the chance level. Consequently, the marginal probability of a random test-taker answering item $j$ correctly is
\begin{align} \label{eq_marg_prob_mass_dist_of_y_population}
    \begin{aligned}
        & \mathrm{P}(U_j = 1\mid\iota_j, \boldsymbol{\xi}_j, g_j) \\
        & = \mathrm{P}(U_j = 1, \Delta_j = 0 \mid \iota_j, g_j) + \mathrm{P}(U_j = 1, \Delta_j = 1 \mid \iota_j, \boldsymbol{\xi}_j).
    \end{aligned}
\end{align}

Third, let $f_{\phi, \tau} = f_{\tau\mid \phi}f_{\phi}$ denote the joint density function of $\phi$ and $\tau$ and $f_{\tau\mid \phi}$ the conditional density function of $\tau$ given $\phi$. For a random, engaged test-taker, the density function of $L_j$ at $l_0$ is
\begin{align} \label{eq_marg_density_of_L_engaged_population_part1}
    \begin{aligned}
    f_{L_j}^1(l_0 \mid \iota_j, \alpha_j, \beta_j) 
    & = \iint_{\mathbb{R}} \mathrm{P}(\Delta_j = 1 \mid  \phi, \iota_j) f_{L_j\mid \Delta_j=1}(l_0; \beta_j - \tau, \alpha_j^2) \\ 
    & \quad \times f_{\tau\mid \phi}(\tau; \mu_{\tau\mid \phi}, \sigma_{\tau\mid \phi}^2) f_{\phi}(\phi; \mu_{\phi}, \sigma_{\phi}^2) d\tau d\phi \\
    & = \int_{\mathbb{R}} \mathrm{P}(\Delta_j = 1 \mid  \phi, \iota_j) f_{\phi}(\phi; \mu_{\phi}, \sigma_{\phi}^2) \\ 
    & \quad \times \int_{\mathbb{R}} f_{L_j\mid \Delta_j=1}(l_0; \beta_j - \tau, \alpha_j^2) f_{\tau\mid \phi}(\tau; \mu_{\tau\mid \phi}, \sigma_{\tau\mid \phi}^2) d\tau d\phi.
    \end{aligned}
\end{align}
In Eq.~\eqref{eq_marg_density_of_L_engaged_population_part1}, the integral with respect to $\tau$ is
\begin{align} \label{eq_int_tau_marg_density_of_L_engaged_population_part1}
    \begin{aligned}
    & \int_{\mathbb{R}} f_{L_j\mid \Delta_j=1}(l_0; \beta_j - \tau, \alpha_j^2) f_{\tau\mid \phi}(\tau; \mu_{\tau\mid \phi}, \sigma_{\tau\mid \phi}^2) d\tau \\
    & = f(\beta_j - l_0; \mu_{\tau\mid \phi}, \alpha_j^2 + \sigma_{\tau\mid \phi}^2) \\
    & = \frac{1}{\sqrt{2\pi(\alpha_j^2+\sigma_{\tau\mid \phi}^2)}}\exp{\left(-\frac{(\beta_j - l_0 - \mu_{\tau\mid \phi})^2}{2(\alpha_j^2+\sigma_{\tau\mid \phi}^2)}\right)},
    \end{aligned}
\end{align}
where, if we plug in $\mu_{\tau\mid \phi} = \mu_{\tau} + \frac{\sigma_{\tau\phi}}{\sigma_\phi^2}(\phi - \mu_\phi)$, the numerator in the exponential function can be expressed as
\begin{align*}
    & (\beta_j - l_0 - \mu_{\tau\mid \phi})^2 = (\beta_j - l_0)^2 - 2(\beta_j - l_0)\mu_{\tau\mid \phi}+\mu_{\tau\mid \phi}^2 \\
    & = (\beta_j - l_0)^2 - 2(\beta_j - l_0)\left[\mu_{\tau} + \frac{\sigma_{\tau\phi}}{\sigma_\phi^2}(\phi - \mu_\phi)\right]+\left[\mu_{\tau} + \frac{\sigma_{\tau\phi}}{\sigma_\phi^2}(\phi - \mu_\phi)\right]^2 \\
    & = (\beta_j - l_0)^2 - 2(\beta_j - l_0)\mu_\tau + \mu_\tau^2 \\
    & \quad - 2(\beta_j - l_0)\frac{\sigma_{\tau\phi}}{\sigma_\phi^2}(\phi - \mu_\phi) + 2\mu_\tau\frac{\sigma_{\tau\phi}}{\sigma_\phi^2}(\phi - \mu_\phi) + \left[\frac{\sigma_{\tau\phi}}{\sigma_\phi^2}(\phi - \mu_\phi)\right]^2 \\
    & = (\beta_j - l_0 - \mu_\tau)^2 - 2(\beta_j - l_0 - \mu_\tau)\frac{\sigma_{\tau\phi}}{\sigma_\phi^2}(\phi - \mu_\phi) + \left[\frac{\sigma_{\tau\phi}}{\sigma_\phi^2}(\phi - \mu_\phi)\right]^2 \\
    & = (\beta_j - l_0 - \mu_\tau)^2 + g(l_0,\phi),
\end{align*}
which means that
\begin{align*}
    & f(\beta_j - l_0; \mu_{\tau\mid \phi}, \alpha_j^2 + \sigma_{\tau\mid \phi}^2) = \frac{1}{\sqrt{2\pi(\alpha_j^2+\sigma_{\tau\mid \phi}^2)}}\exp{\left(-\frac{(\beta_j - l_0 - \mu_{\tau\mid \phi})^2}{2(\alpha_j^2+\sigma_{\tau\mid \phi}^2)}\right)} \\ 
    & = \frac{1}{\sqrt{2\pi(\alpha_j^2+\sigma_{\tau\mid \phi}^2)}}\exp{\left(-\frac{(\beta_j - l_0 - \mu_\tau)^2 + g(l_0,\phi)}{2(\alpha_j^2+\sigma_{\tau\mid \phi}^2)}\right)} \\
    & = \frac{1}{\sqrt{2\pi(\alpha_j^2+\sigma_{\tau\mid \phi}^2)}}\exp{\left(-\frac{(\beta_j - l_0 - \mu_\tau)^2}{2(\alpha_j^2+\sigma_{\tau\mid \phi}^2)}\right)}\exp{\left(-\frac{g(l_0,\phi)}{2(\alpha_j^2+\sigma_{\tau\mid \phi}^2)}\right)} \\
    & = f(l_0; \beta_j-\mu_{\tau}, \alpha_j^2+\sigma_{\tau\mid \phi}^2)\times \exp{\left(-\frac{g(l_0,\phi)}{2(\alpha_j^2+\sigma_{\tau\mid \phi}^2)}\right)}.
\end{align*}
Finally,
\begin{align} \label{eq_marg_density_of_L_engaged_population_part2}
    \begin{aligned}
    & f_{L_j}^1(l_0 \mid \iota_j, \alpha_j, \beta_j) \\ 
    & = \int_{\mathbb{R}} \mathrm{P}(\Delta_j = 1 \mid  \phi, \iota_j) f_{\phi}(\phi; \mu_{\phi}, \sigma_{\phi}^2) f(\beta_j - l_0; \mu_{\tau\mid \phi}, \alpha_j^2 + \sigma_{\tau\mid \phi}^2) d\phi \\
    & = \int_{\mathbb{R}} \mathrm{P}(\Delta_j = 1 \mid  \phi, \iota_j) f_{\phi}(\phi; \mu_{\phi}, \sigma_{\phi}^2) \\
    & \quad \times f(l_0; \beta_j-\mu_{\tau}, \alpha_j^2+\sigma_{\tau\mid \phi}^2)\times \exp{\left(-\frac{g(l_0,\phi)}{2(\alpha_j^2+\sigma_{\tau\mid \phi}^2)}\right)} d\phi \\
    & = f(l_0; \beta_j-\mu_{\tau}, \alpha_j^2+\sigma_{\tau\mid \phi}^2) \\
    & \quad \times \int_{\mathbb{R}} \mathrm{P}(\Delta_j = 1 \mid  \phi, \iota_j) f_{\phi}(\phi; \mu_{\phi}, \sigma_{\phi}^2) \exp{\left(-\frac{g(l_0,\phi)}{2(\alpha_j^2+\sigma_{\tau\mid \phi}^2)}\right)} d\phi,
    \end{aligned}
\end{align}
which is not proportional to a normal density with respect to log RT but a normal density times a function of log RT. Moreover, the probability of a random test-taker answering in the engaged state and having a log RT of less than or equal to $l_0$ is
\begin{align} \label{eq_marg_prob_mass_dist_of_x_engaged_population_part1}
    \begin{aligned}
    & \mathrm{P}(X_{j} = 1, \Delta_j = 1\mid \iota_j, \alpha_j, \beta_j) \\
    & = \iint_{\mathbb{R}} \mathrm{P}(X_j = 1, \Delta_j = 1 \mid  \phi, \tau, \iota_j, \alpha_j, \beta_j) f_{\phi,\tau}(\phi, \tau; \mu_{\phi,\tau}, \Sigma_{\phi,\tau}) d\tau d\phi \\
    & = \iint_{\mathbb{R}} \mathrm{P}(\Delta_j = 1 \mid  \phi, \iota_j) \mathrm{P}(X_j = 1 \mid  \Delta_j = 1, \tau, \alpha_j, \beta_j) \\ 
    & \quad \times f_{\tau\mid \phi}(\tau; \mu_{\tau\mid \phi}, \Sigma_{\tau\mid \phi}) f_{\phi}(\phi; \mu_{\phi}, \sigma_{\phi}^2) d\tau d\phi \\
    & = \int_{\mathbb{R}} \mathrm{P}(\Delta_j = 1 \mid  \phi, \iota_j) f_{\phi}(\phi; \mu_{\phi}, \sigma_{\phi}^2) \\ 
    & \quad \times \int_{\mathbb{R}} \mathrm{P}(X_j = 1 \mid  \Delta_j = 1, \tau, \alpha_j, \beta_j) f_{\tau\mid \phi}(\tau; \mu_{\tau\mid \phi}, \sigma_{\tau\mid \phi}^2) d\tau d\phi.
    \end{aligned}
\end{align}
In Eq.~\eqref{eq_marg_prob_mass_dist_of_x_engaged_population_part1}, the integral with respect to $\tau$ is
\begin{align} \label{eq_int_tau_marg_prob_mass_dist_of_x_engaged_population_part1}
    \begin{aligned}
    &  \int_{\mathbb{R}} \mathrm{P}(X_j = 1 \mid  \Delta_j = 1, \tau, \alpha_j, \beta_j) f_{\tau\mid \phi}(\tau; \mu_{\tau\mid \phi}, \sigma_{\tau\mid \phi}^2) d\tau \\
    & =  \int_{\mathbb{R}} \mathrm{P}(L_j \leq l_0 \mid  \Delta_j = 1, \tau, \alpha_j, \beta_j) f_{\tau\mid \phi}(\tau; \mu_{\tau\mid \phi}, \sigma_{\tau\mid \phi}^2) d\tau \\
    & = \int_{\mathbb{R}} \Phi\left(\frac{l_0 - \beta_j + \tau}{\alpha_j}\right) f_{\tau\mid \phi}(\tau; \mu_{\tau\mid \phi}, \sigma_{\tau\mid \phi}^2) d\tau \\
    & = \Phi\left(\frac{\mu_{\tau\mid \phi} + l_0 - \beta_j}{\sqrt{\alpha_j^2+\sigma_{\tau\mid \phi}^2}}\right),
    \end{aligned}
\end{align}
and thus
\begin{align} \label{eq_marg_prob_mass_dist_of_x_engaged_population_part2}
    \begin{aligned}
    & \mathrm{P}(X_{j} = 1, \Delta_j = 1\mid \iota_j, \alpha_j, \beta_j) \\ 
    & = \int_{\mathbb{R}} \mathrm{P}(\Delta_j = 1 \mid  \phi, \iota_j) f_{\phi}(\phi; \mu_{\phi}, \sigma_{\phi}^2) \Phi\left(\frac{\mu_{\tau\mid \phi} + l_0 - \beta_j}{\sqrt{\alpha_j^2+\sigma_{\tau\mid \phi}^2}}\right) d\phi.
    \end{aligned}
\end{align}
In contrast, for a random, disengaged test-taker, the density function of $L_j$ at $l_0$ does not depend on the speed parameter, and thus 
\begin{align} \label{eq_marg_density_of_L_disengaged_population}
    f_{L_j}^0(l_0 \mid \iota_j, \mu_c, \sigma_c) = \mathrm{P}(\Delta_j = 0)f_{L_j \mid \Delta_j=0}(l_0; \mu_c, \sigma_c^2).
\end{align}
In addition, 
\begin{align} \label{eq_marg_prob_mass_dist_of_x_disengaged_population}
    \mathrm{P}(X_{j} = 1, \Delta_j = 0 \mid \iota_j, \mu_c, \sigma_c) = \mathrm{P}(\Delta_j = 0)\Phi\left(\frac{l_0 - \mu_c}{\sigma_c}\right).
\end{align}
Consequently, for a random test-taker, the density function of $L_j$ at $l_0$ is
\begin{align} \label{eq_marg_density_of_L_population}
    \begin{aligned}
    f_{L_j}(l_0 \mid \iota_j, \alpha_j, \beta_j, \mu_c, \sigma_c) = f_{L_j}^0(l_0 \mid \iota_j, \mu_c, \sigma_c) + f_{L_j}^1(l_0 \mid \iota_j, \alpha_j, \beta_j),
    \end{aligned}
\end{align}
and the marginal probability of a random test-taker having a log RT of less than or equal to $l_0$ is
\begin{align} \label{eq_marg_prob_mass_dist_of_x_population}
    \begin{aligned}
    & \mathrm{P}(X_j = 1 \mid \iota_j, \alpha_j, \beta_j, \mu_c, \sigma_c) \\
    & = \mathrm{P}(X_{j} = 1, \Delta_j = 0 \mid \iota_j, \mu_c, \sigma_c) + \mathrm{P}(X_{j} = 1, \Delta_j = 1 \mid \iota_j, \alpha_j, \beta_j).
    \end{aligned}
\end{align}

Finally, let $f_{\phi, \theta, \tau} = f_{\tau\mid \phi,\theta}f_{\theta\mid \phi}f_{\phi}$ denote the joint density function of $\phi$, $\theta$, and $\tau$ and $f_{\tau\mid \phi,\theta}$ the conditional density function of $\tau$ given $\phi$ and $\theta$. The probability of a random test-taker being in the engaged state, answering correctly, and having a log RT of less than or equal to $l_0$ is
\begin{align} \label{eq_joint_prob_mass_dist_of_x_and_y_engaged_population}
    \begin{aligned}
    & \mathrm{P}(X_j = 1, U_j = 1, \Delta_j = 1\mid \iota_j, \boldsymbol{\xi}_j, \alpha_j, \beta_j) \\
    & = \iiint_{\mathbb{R}} \mathrm{P}(X_j = 1, U_j = 1, \Delta_j = 1 \mid  \phi, \theta, \tau, \iota_j, \boldsymbol{\xi}_j, \alpha_j, \beta_j) \\
    & \quad \times f_{\phi,\theta,\tau}(\phi, \theta, \tau; \mu_{\mathcal{P}}, \Sigma_{\mathcal{P}}) d\tau d\theta d\phi \\
    & = \iiint_{\mathbb{R}} \mathrm{P}(\Delta_j = 1 \mid  \phi, \iota_j) \mathrm{P}(U_j = 1 \mid  \Delta_j = 1, \theta, \boldsymbol{\xi}_j) \\
    & \quad \times \mathrm{P}(X_j = 1 \mid  \Delta_j = 1, \tau, \alpha_j, \beta_j) \\
    & \quad \times f_{\tau\mid \phi,\theta}(\tau; \mu_{\tau\mid \phi,\theta}, \sigma_{\tau\mid \phi,\theta}^2) f_{\theta\mid \phi}(\theta; \mu_{\theta\mid \phi}, \sigma_{\theta\mid \phi}^2) f_{\phi}(\phi; \mu_{\phi}, \sigma_{\phi}^2) d\tau d\theta d\phi \\
    & = \int_{\mathbb{R}} \mathrm{P}(\Delta_j = 1 \mid  \phi, \iota_j) f_{\phi}(\phi; \mu_{\phi}, \sigma_{\phi}^2) \\
    & \quad \times \int_{\mathbb{R}} \mathrm{P}(U_j = 1 \mid  \Delta_j = 1, \theta, \boldsymbol{\xi}_j) f_{\theta\mid \phi}(\theta; \mu_{\theta\mid \phi}, \sigma_{\theta\mid \phi}^2) \\
    & \quad \times \int_{\mathbb{R}} \mathrm{P}(X_j = 1 \mid  \Delta_j = 1, \tau, \alpha_j, \beta_j) f_{\tau\mid \phi,\theta}(\tau; \mu_{\tau\mid \phi,\theta}, \sigma_{\tau\mid \phi,\theta}^2) d\tau d\theta d\phi \\
    & = \int_{\mathbb{R}} \mathrm{P}(\Delta_j = 1 \mid  \phi, \iota_j) f_{\phi}(\phi; \mu_{\phi}, \sigma_{\phi}^2) \\
    & \quad \times \int_{\mathbb{R}} \mathrm{P}(U_j = 1 \mid  \Delta_j = 1, \theta, \boldsymbol{\xi}_j) f_{\theta\mid \phi}(\theta; \mu_{\theta\mid \phi}, \sigma_{\theta\mid \phi}^2) \\
    & \quad \times \Phi\left(\frac{\mu_{\tau\mid \phi,\theta} + l_0 - \beta_j}{\sqrt{\alpha_j^2+\sigma_{\tau\mid \phi,\theta}^2}}\right) d\theta d\phi.
    \end{aligned}
\end{align}
In contrast, the probability of a random test-taker being in the disengaged state, answering correctly, and having a lot RT of less than or equal to $l_0$ is not dependent on the ability and speed parameters, and thus
\begin{align} \label{eq_joint_prob_mass_dist_of_x_and_y_disengaged_population}
    \begin{aligned}
    \mathrm{P}(X_j = 1, U_j = 1, \Delta_j = 0 \mid \iota_j, g_j, \mu_c, \sigma_c) = \mathrm{P}(\Delta_j = 0)\Phi\left(\frac{l_0 - \mu_c}{\sigma_c}\right)g_j.
    \end{aligned}
\end{align}
Consequently, the joint probability of a random test-taker answering correctly and having a log RT less than or equal to $l_0$ is
\begin{align} \label{eq_joint_prob_mass_dist_of_x_and_y_population}
    \begin{aligned}
    & \mathrm{P}(X_j = 1, U_j = 1 \mid \iota_j, \boldsymbol{\xi}_j, g_j, \alpha_j, \beta_j, \mu_c, \sigma_c) \\
    & = \mathrm{P}(X_j = 1, U_j = 1, \Delta_j = 0 \mid \iota_j, g_j, \mu_c, \sigma_c) \\
    & \quad + \mathrm{P}(X_j = 1, U_j = 1, \Delta_j = 1 \mid \iota_j, \boldsymbol{\xi}_j, \alpha_j, \beta_j).
    \end{aligned}
\end{align}

MIRT is computed by plugging the marginal probabilities $\mathrm{P}(U_j = 1)$, $\mathrm{P}(X_j = 1)$, and $\mathrm{P}(X_j = 1, U_j = 1)$, as well as $\mathrm{P}(U_j = 0)$, $\mathrm{P}(X_j = 0)$, $\mathrm{P}(X_j = 0, U_j = 1)$, $\mathrm{P}(X_j = 1, U_j = 0)$, and $\mathrm{P}(X_j = 0, U_j = 0)$, into the formula for MI:
\begin{align} \label{eq_MIRT_def_v2}
    \begin{aligned}
        \mathrm{I}(X_j,U_j) = \mathrm{H}(U_j) - \sum_{x\in\{0,1\}}\mathrm{P}(X_j = x)\mathrm{H}(U_j \mid X_j=x).
    \end{aligned}
\end{align}
We see that, in this case, we could utilise the MaxMI-22 method to identify a log-time threshold to differentiate between RRB and SB.

\section{CUMP and MLN in the ILC-IRT Framework} \label{sec_CUMP_and_MLN_in_ILC-IRT}

The CUMP method~\citep{guo2016cump} is not theoretically justified at the population level in the ILC-IRT framework, because it is based on the noisiness of the CUMP curve around $g_j$, which can only be achieved in finite samples. At the population level, the CUMP curve can be expressed as
\begin{align*}
    f_{\mathrm{CUMP}}(l_0) 
    &\coloneq \mathrm{P}(U_j = 1 \mid X_j = 1, \iota_j, \boldsymbol{\xi}_j, g_j, \alpha_j, \beta_j, \mu_c, \sigma_c) \\ 
    &= \frac{\mathrm{P}(X_j=1,U_j=1 \mid \iota_j, \boldsymbol{\xi}_j, g_j, \alpha_j, \beta_j, \mu_c, \sigma_c)}{\mathrm{P}(X_j=1 \mid \iota_j, \alpha_j, \beta_j, \mu_c, \sigma_c)}.
\end{align*}
In certain conditions, $f_{\mathrm{CUMP}}(l_0) > g_j$ or $f_{\mathrm{CUMP}}(l_0) < g_j$ for every $l_0$. In these cases, the CUMP threshold approaches zero as the sample size approaches infinity at the realised level. For example, if the probability of a correct response for an engaged participant was modelled via the 3PL IRT model and the pseudo-guessing parameter was greater than or equal to the chance level ($c_j \geq g_j$), even the engaged participant with the lowest success rate would beat the chance level, and thus, $f_{\mathrm{CUMP}}(l_0) > g_j$ for every $l_0$. However, in some other framework that, for example, would consider that the left tail of the engaged RT distribution should stop at a certain time point considered too small even for the fastest engaged responses, the response accuracy of the engaged test-takers would not influence the CUMP curve beyond this threshold and the method would work.

 In the MLN~\citep{rios2020mln} method, the estimated RT density does not align with the ILC-IRT framework when the engagement probability is assumed to be dependent on the person. At the population level, the MLN method corresponds to minimizing
\begin{align*}
    f_{\mathrm{MLN}}(l_0) \coloneq \frac{\mathrm{P}(\Delta_j = 0 \mid \iota_j)f_{L_j\mid\Delta_j=0}(l_0; \mu_0, \sigma_0^2) + \mathrm{P}(\Delta_j = 1 \mid \iota_j)f_{L_j\mid\Delta_j=1}(l_0; \mu_1,\sigma_1^2)}{\exp{(l_0)}}
\end{align*}
with respect to $l_0\in\{\mu_0-\sigma_0^2,\mu_1-\sigma_1^2\}$, where the log RT density of the engaged component is assumed to be proportional to a normal density. However, previously we showed that due to the effect of $\phi$, this is not true.

\section{Misclassification Rate of the Log RT Indicator} \label{sec_MR}

The log-time point $l_0$ can be interpreted as the log-time threshold for identifying disengaged responses, and thus the complement of $X_j$ is an estimator of $\Delta_j$. Therefore, the misclassification rate of $X_j$ can be written as a function of $l_0$ as the probability that $X_j$ equals $\Delta_j$, that is,
\begin{align} \label{eq_misclas_rate_def}
    \begin{aligned}
     \mathrm{MR}_{X_j}(l_0) 
     &\coloneq \mathrm{P}(X_j = \Delta_j\mid \iota_j,\alpha_j,\beta_j) \\
     &= \mathrm{P}(X_j = 1, \Delta_j = 1\mid \iota_j,\alpha_j,\beta_j) + \mathrm{P}(X_j = 0, \Delta_j = 0\mid \iota_j,\mu_c,\sigma_c) \\
     &= \int_{\mathbb{R}} \mathrm{P}(\Delta_j = 1 \mid  \phi, \iota_j) f_{\phi}(\phi; \mu_{\phi}, \sigma_{\phi}^2) \Phi\left(\frac{\mu_{\tau\mid \phi} + l_0 - \beta_j}{\sqrt{\alpha_j^2+\sigma_{\tau\mid \phi}^2}}\right) d\phi \\
     & \quad + \mathrm{P}(\Delta_j = 0)\left(1-\Phi\left(\frac{l_0 - \mu_c}{\sigma_c}\right)\right),
    \end{aligned}
\end{align}
where the first term is the probability that a true engaged response is classified as a disengaged response (false positive) and the second term is the probability that a true disengaged response is classified as an engaged response (false negative). The MR has both a global minimum and a global maximum. It is minimized or maximized at a time point where its first derivative is equal to zero. The first derivative of $\mathrm{MR}_{X_j}(l_0)$ is
\begin{equation*}
    \frac{d\mathrm{MR}_{X_j}(l_0)}{dl_0} = \frac{d\mathrm{P}(X_j = 1, \Delta_j = 1\mid \iota_j,\alpha_j,\beta_j)}{dl_0} + \frac{d\mathrm{P}(X_j = 0, \Delta_j = 0\mid \iota_j,\mu_c,\sigma_c)}{dl_0}.
\end{equation*}
Due to the integrals in both terms, which cannot be solved analytically, we need numerical methods to find the zeros of the first derivative of $\mathrm{MR}_{X_j}(l_0)$.

However, if $\mathrm{P}(\Delta_j = 1 \mid \lambda_j) = \lambda_j$ was independent of the person, we would have
\begin{align*}
    \mathrm{P}(X_j = 1, \Delta_j = 1 \mid  \lambda_j, \alpha_j, \beta_j) 
    &= \mathrm{P}(\Delta_j = 1 \mid \lambda_j)\mathrm{P}(X_j = 1 \mid  \Delta_j = 1, \alpha_j, \beta_j) \\
    &= \lambda_j\int_{\mathbb{R}}\mathrm{P}(X_j = 1 \mid  \Delta_j = 1, \tau,\alpha_j,\beta_j)f_{\tau}(\tau; \mu_{\tau}, \sigma_{\tau}^2)d\tau \\
    &= \lambda_j \Phi\left(\frac{\mu_{\tau} + l_0 - \beta_j}{\sqrt{\alpha_j^2+\sigma_{\tau}^2}}\right)
\end{align*}
and
\begin{align*}
    \mathrm{P}(X_j = 0, \Delta_j = 0\mid \lambda_j, \mu_c, \sigma_c)
    &= \mathrm{P}(\Delta_j = 0 \mid \lambda_j)\mathrm{P}(X_j = 0 \mid  \Delta_j = 0,\mu_c, \sigma_c) \\
    &= (1-\lambda_j)\left(1-\Phi\left(\frac{l_0 - \mu_c}{\sigma_c}\right)\right),
\end{align*}
and the derivative would be
\begin{align*}
\frac{d\mathrm{MR}_{X_j}(l_0)}{dl_0} 
&= \frac{d\left[\lambda_j\Phi\left(\frac{\mu_{\tau} + l_0 - \beta_j}{\sqrt{\alpha_j^2+\sigma_{\tau}^2}}\right) + (1-\lambda_j)\left(1-\Phi\left(\frac{l_0 - \mu_c} {\sigma_c}\right)\right)\right]}{dl_0} \\
&= \frac{\lambda_j}{\sqrt{\alpha_j^2+\sigma_{\tau}^2}}h\left(\frac{\mu_{\tau} + l_0 - \beta_j}{\sqrt{\alpha_j^2+\sigma_{\tau}^2}}\right) - \frac{1-\lambda_j}{\sigma_c}h\left(\frac{l_0-\mu_c}{\sigma_c}\right).
\end{align*}
Let $\mu_1 = \beta_j - \mu_{\tau}$, $\mu_2 = \mu_c$, $\sigma_1 = \sqrt{\alpha_j^2+\sigma_{\tau}^2}$, and $\sigma_c = \sigma_2$. We can find the zeros of the derivative as follows:
\begin{align*}
& \frac{d\mathrm{MR}_{X_j}(l_0)}{dl_0} = \frac{\lambda_j}{\sigma_1}h\left(\frac{l_0-\mu_1}{\sigma_1}\right) - \frac{1-\lambda_j}{\sigma_2}h\left(\frac{l_0-\mu_2}{\sigma_2}\right) = 0 \\
& \Leftrightarrow \frac{\lambda_j}{\sigma_1}h\left(\frac{l_0-\mu_1}{\sigma_1}\right) = \frac{1-\lambda_j}{\sigma_2}h\left(\frac{l_0-\mu_2}{\sigma_2}\right) \\
& \Leftrightarrow \frac{\lambda_j}{\sigma_1}\cdot\frac{1}{\sqrt{2\pi}}\exp\left(-\frac{(l_0-\mu_1)^2}{2\sigma_1^2}\right) = \frac{1-\lambda_j}{\sigma_2}\cdot\frac{1}{\sqrt{2\pi}}\exp\left(-\frac{(l_0-\mu_2)^2}{2\sigma_2^2}\right) \\
& \Leftrightarrow \ln\left(\frac{\lambda_j}{\sigma_1}\right) - \frac{(l_0-\mu_1)^2}{2\sigma_1^2} = \ln\left(\frac{1-\lambda_j}{\sigma_2}\right) - \frac{(l_0-\mu_2)^2}{2\sigma_2^2} \\
& \Leftrightarrow \ln\left(\frac{\lambda_j}{\sigma_1}\right) - \ln\left(\frac{1-\lambda_j}{\sigma_2}\right) = \frac{(l_0-\mu_1)^2}{2\sigma_1^2} - \frac{(l_0-\mu_2)^2}{2\sigma_2^2} \\
& \Leftrightarrow 2\sigma_2^2\sigma_1^2\left[\ln\left(\frac{\lambda_j}{\sigma_1}\right) - \ln\left(\frac{1-\lambda_j}{\sigma_2}\right)\right] = \sigma_2^2(l_0-\mu_1)^2 - \sigma_1^2(l_0-\mu_2)^2 \\
& \Leftrightarrow \left(\sigma_2^2 - \sigma_1^2\right)l_0^2 + 2\left(\sigma_1^2\mu_2 - \sigma_2^2\mu_1\right)l_0 \\ 
&+ \left(\sigma_2^2\mu_1^2 - \sigma_1^2\mu_2^2 - 2\sigma_2^2\sigma_1^2\left[\ln\left(\frac{\lambda_j}{\sigma_1}\right) - \ln\left(\frac{1-\lambda_j}{\sigma_2}\right)\right]\right) = 0,
\end{align*}
which is a quadratic equation of the form $Al_0^2+Bl_0+C=0$, where $A = \sigma_2^2 - \sigma_1^2$, $B = 2\left(\sigma_1^2\mu_2 - \sigma_2^2\mu_1\right)$, and $C = \sigma_2^2\mu_1^2 - \sigma_1^2\mu_2^2 - 2\sigma_2^2\sigma_1^2\left[\ln\left(\frac{\lambda_j}{\sigma_1}\right) - \ln\left(\frac{1-\lambda_j}{\sigma_2}\right)\right]$. Therefore, the zeros are
\begin{equation*}
\tilde{l}_1 = \frac{-B-\sqrt{B^2-4AC}}{2A} \ \mathrm{and} \ \tilde{l}_2 = \frac{-B+\sqrt{B^2-4AC}}{2A}.
\end{equation*}
Notably, if we denote the log RT density functions of the components as $f_{L_j\mid \Delta_j=1} = f_{1}(l_0;\mu_1,\sigma_1^2)$ and $f_{L_j\mid \Delta_j=0} = f_{2}(l_0;\mu_2,\sigma_2^2)$, it applies that
\begin{align*}
\lambda_jf_{L_j\mid \Delta_j=1} = \lambda_jf_{1}(l_0;\mu_1,\sigma_1^2) = \frac{\lambda_j}{\sigma_1}h\left(\frac{l_0-\mu_1}{\sigma_1}\right) \ \mathrm{and} \\
(1-\lambda_j)f_{L_j\mid \Delta_j=0} = (1-\lambda_j)f_{2}(l_0;\mu_2,\sigma_2^2) = \frac{1-\lambda_j}{\sigma_2}h\left(\frac{l_0-\mu_2}{\sigma_2}\right).
\end{align*}
We see that $d\mathrm{MR}_{X_j}(l_0)/dl_0 = \lambda_jf_{L_j\mid \Delta_j=1} - (1-\lambda_j)f_{L_j\mid \Delta_j=0}$. Therefore, the zeros of $d\mathrm{MR}_{X_j}(l_0)/dl_0$ are also the zeros of the difference between $\lambda_jf_{L_j\mid \Delta_j=1}$ and $(1-\lambda_j)f_{L_j\mid \Delta_j=0}$, and the time point where $\mathrm{MR}_{X_j}(l_0)$ is minimized or maximized is also the crossing point of the functions $\lambda_jf_{L_j\mid \Delta_j=1}$ and $(1-\lambda_j)f_{L_j\mid \Delta_j=0}$.

To have an understanding of the shape of the misclassification rate as a function of log RT in realistic situations, we need to make some assumptions about the relative magnitudes of the parameters. First, \citet{rios2022meta} found an average RRB proportion of 0.1 in low-stakes assessments. Therefore, $1-\lambda_j < 0.5$ should be a realistic assumption. Second, according to our conceptual understanding of RRB and SB, we assume $\mu_1 > \mu_2$. Finally, previous research has found larger variance for the log RTs in the RBB component than in the SB component~\citep[e.g.,][]{holopainen2026,nagy2022multilevel,schnipke1997modeling,ulitzsch2020hierarchical}. Therefore, we assume $\sigma_1 < \sigma_2$. With these assumptions, $d\mathrm{MR}_{X_j}(l_0)/dl_0 < 0$ when $l_0 < \tilde{l}_1$, $d\mathrm{MR}_{X_j}(l_0)/dl_0 > 0$ when $\tilde{l}_1 < l_0 < \tilde{l}_2$, and $d\mathrm{MR}_{X_j}(l_0)/dl_0 < 0$ when $l_0 > \tilde{l}_2$. Therefore, the minimum of $\mathrm{MR}(l_0)$ is located at the smaller zero of $d\mathrm{MR}_{X_j}(l_0)/dl_0$.

In addition, the minimum of $\mathrm{MR}(l_0)$ is located at a time point smaller than $\mu_1$. This can be easily seen if we substitute $\mu_1$ to the quadratic function $g(x) = Ax^2 + Bx + C$, which represents a parabola that opens to the top:
\begin{align*}
g(\mu_1) 
&= A\mu_1^2+B\mu_1+C = (\sigma_2^2 - \sigma_1^2)\mu_1^2 + 2\left(\sigma_1^2\mu_2 - \sigma_2^2\mu_1\right)\mu_1+C \\
&= \sigma_2^2\mu_1^2 - \sigma_1^2\mu_1^2 + 2\sigma_1^2\mu_2\mu_1 - 2\sigma_2^2\mu_1^2\ + C \\
&= -\sigma_2^2\mu_1^2 - \sigma_1^2\mu_1^2 + 2\sigma_1^2\mu_2\mu_1 + C \\
&= -\sigma_2^2\mu_1^2 - \sigma_1^2\mu_1^2 + 2\sigma_1^2\mu_2\mu_1 + \sigma_2^2\mu_1^2 - \sigma_1^2\mu_2^2 \\
&- 2\sigma_2^2\sigma_1^2\left[\ln\left(\frac{\lambda_j}{\sigma_1}\right) - \ln\left(\frac{1-\lambda_j}{\sigma_2}\right)\right] \\
&= - \sigma_1^2\mu_1^2 + 2\sigma_1^2\mu_2\mu_1 - \sigma_1^2\mu_2^2 - 2\sigma_2^2\sigma_1^2\left[\ln\left(\frac{\sigma_2}{\sigma_1}\right) + \ln\left(\frac{\lambda_j}{1-\lambda_j}\right)\right] \\
&= -\sigma_1^2\left(\mu_1^2 - 2\mu_2\mu_1+\mu_2^2\right) - 2\sigma_2^2\sigma_1^2\left[\ln\left(\frac{\sigma_2}{\sigma_1}\right) + \ln\left(\frac{\lambda_j}{1-\lambda_j}\right)\right] \\
&= -\sigma_1^2(\mu_1 - \mu_2)^2 - 2\sigma_2^2\sigma_1^2\left[\ln\left(\frac{\sigma_2}{\sigma_1}\right) + \ln\left(\frac{\lambda_j}{1-\lambda_j}\right)\right],
\end{align*}
which is negative, because $\sigma_1^2(\mu_1 - \mu_2)^2 > 0$, $\sigma_2^2\sigma_1^2 > 0$, $\ln\left(\frac{\sigma_2}{\sigma_1}\right) > 0$, and $\ln\left(\frac{\lambda_j}{1-\lambda_j}\right) > 0$. Therefore, $\mu_1$ is located between the zeros of $d\mathrm{MR}_{X_j}(l_0)/dl_0$. Note that $\mu_2$ may also be located between the zeros.

\clearpage

\bibliographystyle{apalike}
\bibliography{references}
